\documentclass[%
 reprint,
 amsmath,amssymb,
 aps,
]{revtex4-2}

\usepackage{graphicx}
\usepackage{dcolumn}
\usepackage{bm}
\usepackage{xcolor}
\usepackage{comment}
\usepackage{csquotes}
\usepackage{afterpage}

\DeclareUnicodeCharacter{2243}{\ensuremath{\simeq}}
\DeclareUnicodeCharacter{2299}{\ensuremath{\odot}}
\DeclareUnicodeCharacter{2013}{\textendash}
\DeclareUnicodeCharacter{03B1}{\ensuremath{\alpha}}
\DeclareUnicodeCharacter{039B}{\ensuremath{\Lambda}}

\renewcommand{\selectlanguage}[1]{}

\begin{document}

\preprint{APS/123-QED}

\title{FORGE'D IN THE EARLY UNIVERSE II: CONSTRAINING THE POP III INITIAL MASS FUNCTION}

\author{Yasmine J. Meziani}
\email{ymeziani@caltech.edu}
\affiliation{TAPIR, California Institute of Technology, Mailcode 350-17, Pasadena, CA 91125, USA}

\author{Philip F. Hopkins}
\affiliation{TAPIR, California Institute of Technology, Mailcode 350-17, Pasadena, CA 91125, USA}

\author{Pratik J. Gandhi}
\affiliation{Department of Astronomy, Yale University, New Haven, CT 06520, USA}

\author{Claude-André Faucher-Giguère}
\affiliation{CIERA and Department of Physics and Astronomy, Northwestern University, Evanston, IL 60201, USA}

\author{Michael Y. Grudić}
\affiliation{Center for Computational Astrophysics, Flatiron Institute, 162 5th Ave, New York, NY 10010, USA}

\author{Shivan Khullar}
\affiliation{Center for Computational Astrophysics, Flatiron Institute, 162 5th Ave, New York, NY 10010, USA}


\author{Vasilii Pustovoit}
\affiliation{Canadian Institute for Theoretical Astrophysics, University of Toronto, 60 St. George St., Toronto, ON M5S 3H8, Canada}

\date{\today}

\begin{abstract}

We present a cosmological radiation-magnetohydrodynamics (RMHD) zoom-in simulation of $56$ Population III stars using FORGE'd in FIRE. Our simulation combines a cosmological initial condition centered on an isolated ultra-faint galaxy ($M_{\rm halo} \sim 10^9\,\rm{M_\odot}$ at $z=0$) with individual-star formation and feedback physics, reaching a maximum mass resolution of $\sim 10^{-2}\;\rm{M_\odot}$ and spatial resolution of $\sim 10^{-3}\;\rm{pc}$. We follow continued star formation for just over $500$ Myr, from $z \sim 16$ to $z \sim 7$, substantially longer than what has been achieved before, and track star formation across five distinct minihalos. From our stellar population, we self-consistently resolve an emergent stellar initial mass function (IMF) and find $\alpha = 1.9$ for $dN/dM \propto M^{-\alpha}$ at $\rm M_{\star} \gtrsim 2.2 \, M_{\odot}$. Additionally, we form both single and multiple stellar systems, including binaries and higher-order multiples, as well as stars with masses below $1 \; \rm M_{\odot}$. At early times, the stellar mass distribution is more top-heavy, with a high-mass slope of $\alpha\sim1.4-1.5$, similar to previous Pop. III simulations, and steepens as star formation continues. This suggests that differences between our final IMF and previous studies may partly reflect the much longer evolution followed here, approximately five orders of magnitude longer than in several previous Pop.~III sink-particle simulations. The first star-forming minihalo, which hosts the most massive stars in our simulation, has a shallower IMF slope of $\alpha\approx 1$, while later-forming minihalos produce lower maximum stellar masses in lower-density parent clouds.
\end{abstract}

\maketitle


\section{\label{sec:intro}Introduction}
The initial mass function (IMF) of the first generation of stars, or Population III (Pop.~III) stars, remains one of the central uncertainties in our understanding of early star formation and galaxy evolution. The IMF determines the relative numbers of low- and high-mass stars, which in turn influence supernova explosion rates, stellar luminosities, and the rates and yields of elemental enrichment. Because Pop.~III stars formed from metal-free gas, their formation physics differs substantially from that of present-day star formation, in part because cooling is regulated primarily by molecular hydrogen rather than metals or dust \citep{bromm_first_2004, glover_uncertainties_2008}. Determining the Pop.~III IMF is therefore critical for understanding the first episodes of stellar feedback, chemical enrichment, and the formation of early galaxies.

Early theoretical models and numerical simulations suggested that Pop.~III stars formed in isolation and reached very large stellar masses $\sim 1000 \; M_{\odot}$ \citep{abel_formation_2002, bromm_first_2004, yoshida_era_2004}.
These simulations followed primordial gas collapse to the formation of a protostar but generally did not follow the subsequent evolution of circumstellar disks and long-term protostellar accretion. This early picture therefore suggested a Pop.~III IMF that was top-heavy relative to the present-day ($z=0$) IMF \citep{bromm_first_2004}.

Subsequent simulations with higher resolution and improved physics have revised this picture, with studies that resolve disk formation and fragmentation showing that primordial gas can produce multiple protostars within a single halo \citep{stacy_first_2010, clark_formation_2011, hosokawa_protostellar_2011, greif_simulations_2011, stacy_constraining_2013, hirano_formation_2017}. 
These results suggest that the Pop.~III IMF may span a broader mass range than previously thought and that multiplicity may be common in primordial star-forming systems. 

The binary statistics of Pop.~III stars remain uncertain, but simulations generally find that disk fragmentation can produce small-N multiple systems with dynamically evolving orbital configurations \citep{stacy_first_2010, greif_formation_2012, machida_formation_2013, stacy_constraining_2013}. The resulting multiplicity depends sensitively on accretion physics, radiative feedback, and magnetic fields, all of which influence fragmentation and migration within circumstellar disks. 

Observational constraints on the Pop.~III IMF remain indirect. While no confirmed metal-free stars have been identified in the Milky Way (MW), searches for extremely metal-poor stars in the MW's stellar halo provide limits on the low-mass end of the Pop.~III IMF \citep{beers_discovery_2005, frebel_near-field_2015, magg_observational_2019}. 

Recent observations with the James Webb Space Telescope (JWST) have reported tentative candidates for Pop.~III-dominated galaxies or star clusters \citep{maiolino_small_2024, fujimoto_glimpse_2025, rusta_pristine_2026}. However, the constraints on the metallicities of these systems remain fairly open-ended and do not conclusively distinguish Pop.~III stellar populations from very metal-poor Pop.~II populations. The ionizing spectra and nebular emission from these systems depend strongly on the underlying stellar mass distribution, particularly the abundance of massive stars. These observations therefore provide a way to begin constraining whether the Pop.~III IMF was strongly top-heavy. Because the most massive Pop.~III stars produce much of the hard ionizing radiation, emission-line ratios such as HeII/H$\gamma$ are particularly sensitive to the high-mass end of the IMF and can constrain how top-heavy the stellar population is \citep{schaerer_transition_2003, rusta_pristine_2026}.

Recent advances in cosmological simulations have made it possible to follow Pop.~III star formation with increasingly detailed treatments of protostellar accretion and disk fragmentation, ionizing radiation feedback, and the effects of turbulence and magnetic fields on gravitational collapse and stellar growth \citep[e.g.,][]{stacy_building_2016, sharda_population_2025, sharda_magnetic_2025}. However, most previous work uses a combination of physics and numerical methods known not to produce realistic observed IMFs or stellar multiplicity for stars in nearby metal-rich star-forming regions in the MW at $z=0$. In particular, these simulations generally lack the full combination of protostellar jets and winds, radiative feedback, magnetic fields, and non-equilibrium thermochemistry. The simulation we present in this work extrapolates the predictive power of STARFORGE simulations to unobserved Pop.~III systems while including a realistic cosmological environment simulated with FIRE physics on larger scales \citep{grudic_dynamics_2022, guszejnov_effects_2022, guszejnov_effects_2023}.

In a previous companion paper, we presented a cosmological RMHD simulation that followed the formation and evolution of a single metal-free Pop.~III (proto)star using FORGE'd in FIRE. In this paper, we present a follow-up study to \citet{meziani_forged_2026}, following the growth and evolution of a Pop.~III system containing dozens of stars using FORGE'd in FIRE at around $z \sim 16$, within a fully cosmological zoom-in setup. The simulation spans approximately nine orders of magnitude in spatial scale, from Mpc scales down to $10^{-3} \; \mathrm pc$, reaching a gas mass resolution of $\sim 10^{-2}\;M_\odot$ in the refined regions. We evolve the simulation for more than $500$ Myr, following the long-term evolution of each individual star, with most evolving through the main sequence and the most massive stars eventually undergoing supernovae.

The structure of the paper is as follows. In Section \ref{sec:simulations}, we describe the simulation framework. In Section \ref{sec:results}, we discuss the results, including the system evolution, binary statistics, ejection rate, and initial mass function. We discuss the conclusions and directions for future work in Section \ref{sec:conclusions}.

\section{\label{sec:simulations}Simulations}

\begin{figure*}[p]
    \centering
    \includegraphics[width=\textwidth,height=0.90 \textheight,keepaspectratio]{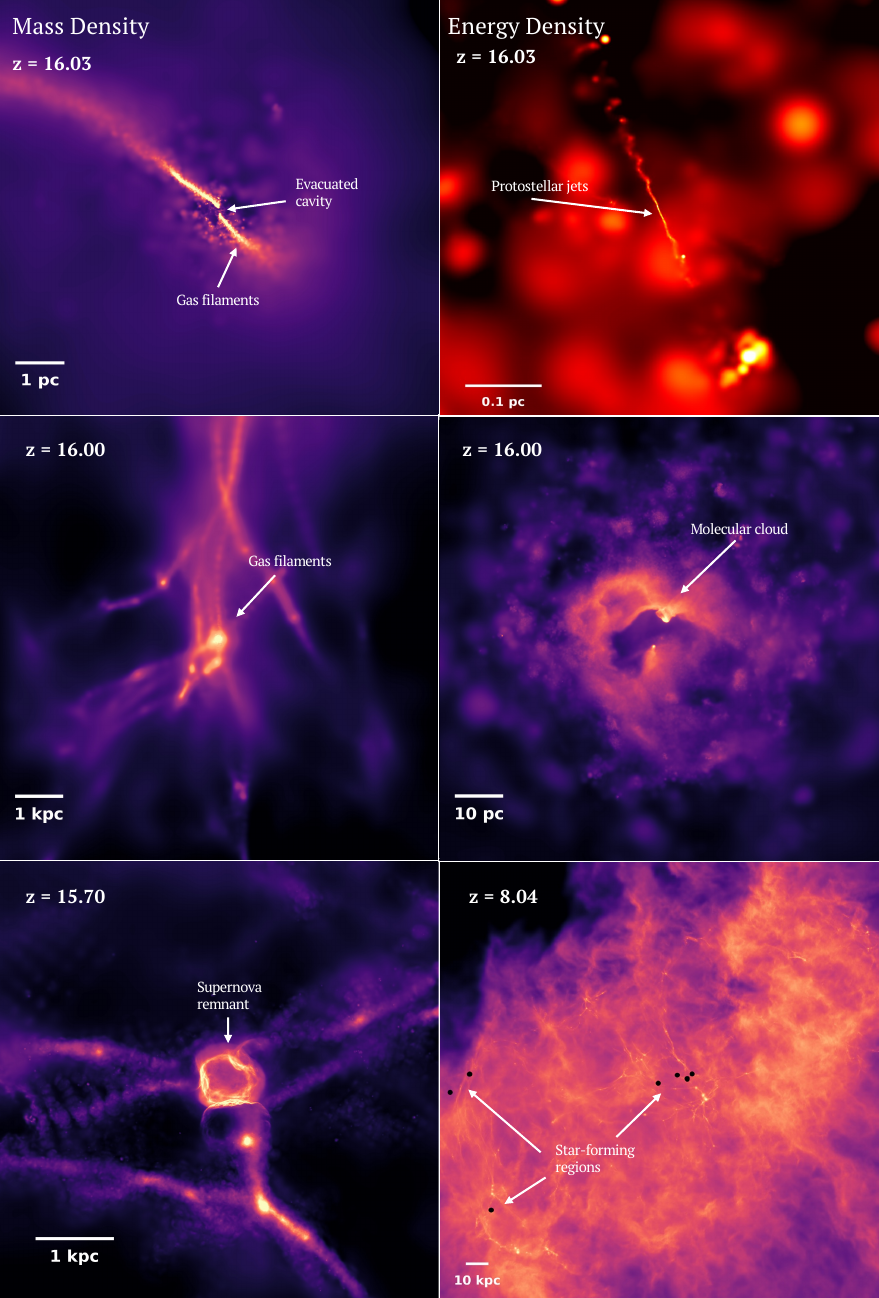}
    \caption{Visualizations of the gas mass and gas kinetic energy surface densities in the cosmological simulation at different redshifts throughout its evolution. Panels shown in purple represent the gas mass surface density, while the top-right panel in red indicates the kinetic energy surface density. Darker colors are regions of lower density, while brighter colors indicate higher density. Multiple panels are used to illustrate different stages of the simulation and to capture its wide dynamic range. The panels show selected regions of the simulation at different spatial scales and evolutionary times. The top-left panel shows a dense gas filament in which stars have formed, with feedback from the newly formed stars evacuating the surrounding gas and producing an evacuated cavity. Star formation occurs along dense gas filaments and within molecular clouds. Protostellar jets are visible before the stars reach the main sequence (top right). The middle-right panel shows a molecular cloud containing filamentary structures on a larger spatial scale. At later times, as the stars evolve and undergo supernovae, supernova remnants appear in the simulation (bottom left). The dots in the bottom-right panel indicate the positions of sink particles.}
    \label{fig:visualizations}
\end{figure*}

\begin{figure*}[t]
    \centering
    \includegraphics[width=\textwidth,height=0.90 \textheight,keepaspectratio]{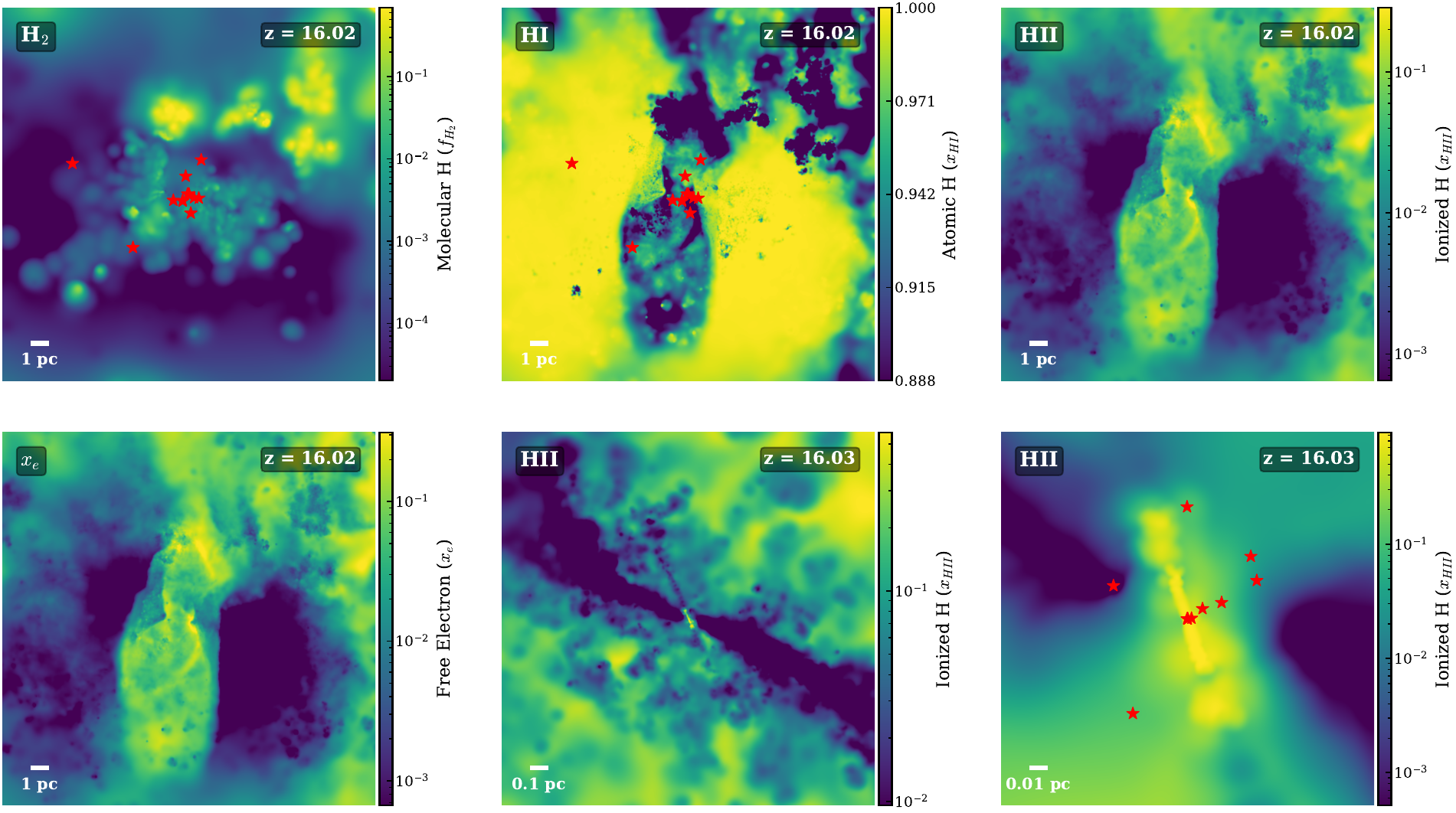}
    \caption{Gas chemistry and ionized structure around a Pop.~III star-forming region at $z\sim16$. \textit{Top row:} column-averaged fractions of molecular hydrogen (H$_2$), atomic hydrogen (HI), and ionized hydrogen (HII). \textit{Bottom left:} column-averaged free-electron fraction, $x_e$. The column-averaged fractions are computed by projecting gas within a spherical region of radius $3r_{\rm max}$ around the central sink particle, where $r_{\max}$ is the half-width of the displayed map measured from the central sink. \textit{Bottom-middle and bottom-right:} progressively smaller-scale views of the HII structure around the (proto)star, shown on $0.1$ pc and $0.01$ pc scales. Red stars mark the sink particle positions in the top-left, top-middle, and bottom-right panels. Each panel is labeled with its redshift. The color scales are logarithmic, with brighter regions indicating higher abundances and independently chosen ranges for each panel.}
    \label{fig:chemistry visualizations}
\end{figure*}

The simulation presented in this paper uses the same cosmological initial conditions and physics as those presented in Paper I \citep{meziani_forged_2026}. We therefore summarize the most relevant aspects here and refer the reader to Paper I for a complete description of the numerical methods and physics implementation. 

The cosmological simulation follows FORGE'd in FIRE \citep{hopkins_forged_2023, hopkins_forged_2024-1, hopkins_forged_2024}, adopting the \textbf{m09} initial condition \citep{hopkins_galaxies_2014, wheeler_sweating_2015, wheeler_be_2019}. The simulation is run with \texttt{GIZMO} \citep{hopkins_new_2015, hopkins_new_2017}, which evolves the coupled equations of gravity, magneto-hydrodynamics, and radiation transport, using the meshless finite-mass (MFM) method \citep{hopkins_accurate_2016, hopkins_anisotropic_2017}. \texttt{GIZMO} evolves gravity using adaptive force softening for gas, fixed softenings for collisionless particles, and a fifth-order Hermite integrator for close gravitational encounters such as binaries \citep{grudic_starforge_2021}. The simulation is initialized with trace seed cosmological magnetic fields of $\sim 10^{-15} \; \rm G$, which are amplified using the constrained gradient MHD solver described in \cite{hopkins_constrained-gradient_2016}. 

The simulation includes a wide range of physics, including self-gravity, magneto-hydrodynamics, radiation transport, and detailed non-equilibrium thermo-chemistry. We also include a spatially uniform, redshift-dependent cosmological UV/X-ray background following \citet{faucher-giguere_cosmic_2020}. Cooling and chemistry are followed over a wide range of densities, temperatures, and metallicities ($n \sim 10^{-8} - 10^{24}\,{\rm cm^{-3}}$, $T \sim 1 - 10^{10}\,{\rm K}$, and $Z = 0 - 10\,Z_{\odot}$), including atomic and molecular processes relevant for primordial gas. The chemical network follows the abundance evolution of ionized, partially ionized, and neutral hydrogen and helium, as well as molecular species including H$_2$ and HD. Sink particles form to represent individual (proto)stars in dense ($n_{\rm H} \gtrsim 10^{10} \; \rm cm^{-3}$), self-gravitating, collapsing regions of gas, following the STARFORGE model \citep{grudic_starforge_2021}. The sink gravitational-softening length ranges from approximately $5$ to $10\,\rm au$ between $z\sim16$ and $z\sim7$.

STARFORGE also includes a mechanism to suppress spurious sink particles, which can arise when unresolved density fluctuations are numerically identified as separate sinks \citep{grudic_starforge_2021}. If extremely low-mass sinks form nearly simultaneously within the same interaction kernel, they can be merged into a single sink particle. This is important for the IMF and multiplicity statistics because spurious sinks could artificially increase the number of low-mass stars.

In the standard STARFORGE stellar evolution model, sink particles progress through stages 0--5, with stage 5 corresponding to the main sequence. In this work, we additionally introduce stellar stages 6 and 7 to identify stars that have reached the end of their stellar lifetimes. This remnant classification is introduced here for the first time and has not been described in previous work. In the current implementation, sink particles with masses $1.4 \leq M_{\rm sink} \leq 20\,M_{\odot}$ undergo supernova explosions at the end of the main sequence and are assigned stage 6. Sink particles with masses $M_{\rm sink} < 1.4\,M_{\odot}$ or $M_{\rm sink} > 20\,M_{\odot}$ are assigned stage 7 and do not undergo supernovae and are meant to represent white dwarfs, black holes, or neutron stars. These stages are used as bookkeeping labels only, and remnant evolution and feedback are not modeled. Unless otherwise stated, these objects are included in the analysis but do not contribute further feedback.

These sink particles evolve self-consistently as they accrete mass, radiate, and launch feedback in the form of protostellar jets and winds \citep{grudic_starforge_2021}. Massive stars eventually end their lives as supernovae. In our simulation, sink particles with masses between $1.4$ and $20\,M_{\odot}$ undergo supernova explosions at the end of their main-sequence lifetimes. Each explosion injects $E_{\rm SN} = 10^{51}\,{\rm erg}$ of energy and ejects the stellar mass isotropically with a characteristic velocity of $v_{\rm SN} \sim 3200\,{\rm km\,s^{-1}}$. The full supernova ejecta energy is injected separately for each individual event, with the ejecta deposited by spawning new gas resolution elements (see \citet{grudic_starforge_2021}). The ejecta carry metal yields following IMF-averaged abundance ratios from \citet{nomoto_nucleosynthesis_2006}. Once a star undergoes a supernova explosion, it ceases all other forms of stellar feedback. In total, $16$ supernovae occur over the $\approx 500$ Myr we evolve this simulation, from $z\sim16$ to $z\sim7$.

Our initial condition is identical to that described in Paper I, centered on the halo \enquote{\textbf{m09}}, an ultra-faint dwarf galaxy at $z=0$ taken from the FIRE cosmological simulation suite \citep{hopkins_galaxies_2014, wheeler_sweating_2015, wheeler_be_2019}. The central dark matter halo has a present-day mass of $M_{\rm halo} \sim 2 \times 10^{9}\,M_{\odot}$ when evolved to $z=0$, and forms its first star at $z \sim 16$. At the time when the first star forms, the halo mass of the largest progenitor halo is $M_{\rm halo} \sim 2.4 \times 10^{6}\; \rm M_{\odot}$.

The hydrodynamical resolution refinement techniques build on those described in Paper I. In this simulation, we refine particles when their mass exceeds a multiple of $0.001$ of the local Jeans mass, ensuring that Jeans-unstable gas is always resolved with at least 1000 cells, consistent with the resolution criterion discussed in \citep{truelove_jeans_1997}. Additional refinement is applied near sink particles so that the gas surrounding the (proto)stars maintains high resolution. This refinement criterion is applied throughout the zoom-in region wherever gas becomes sufficiently dense to collapse and form sink particles, with additional refinement maintained around the resulting sink particles. Recent numerical tests of the GIZMO particle-splitting method find no evidence of artificial fragmentation, with differences in star formation outcomes generally comparable to those arising from stochastic variations in the initial turbulent conditions \citep{khullar_star_2026}.

With this refinement scheme, gas near the (proto)star reaches a typical mass resolution of $\sim 10^{-2} \; \rm M_{\odot}$ with a corresponding spatial resolution of $\sim 10^{-3} \; \rm pc$. This is an order of magnitude lower in resolution than the simulation presented in Paper I \citep{meziani_forged_2026}, where the mass resolution was $\sim 10^{-3} \; \rm M_{\odot}$ and the spatial resolution was $\sim 10^{-4} \; \rm pc$. The lower resolution used in this simulation allows us to evolve the Pop. III stellar population for longer; running the simulation for $\sim 500$ Myr would not be computationally feasible at the resolution used in Paper I. While the lower resolution introduces uncertainties in the low-mass IMF, its turnover, and the multiplicity statistics, previous STARFORGE resolution studies suggest that the high-mass IMF is relatively well converged at our adopted mass resolution of $\sim 10^{-2}\;{\rm M_{\odot}}$ \citep{bate_modelling_1995, grudic_starforge_2021, guszejnov_effects_2022}. However, these convergence tests were performed in different star-forming environments rather than primordial gas. In particular, \citet{prole_fragmentation-induced_2022} found that the number of sink particles in Pop.~III simulations continued to increase with increasing numerical resolution, without reaching convergence.


\subsection{Dark Matter Minihalo Identification} \label{subsec:dark matter minihalo identification}

To identify the dark matter minihalo associated with each star-forming group, we use the gravitational potential and particle positions stored in the simulation snapshots. For each newly formed sink particle (proto-star), we search within $200$ physical pc of its position and identify the dark matter particle with the lowest gravitational potential as the local potential minimum. We then select the $64$ lowest-potential dark matter particles within this region and use their particle IDs solely to identify and track the associated minihalo, rather than to determine its total mass. These particles are used to track the central region of each minihalo, while the halo mass is determined independently from the total dark matter mass enclosed within $R_{200\mathrm{m}}$.

To determine whether stars forming at different times belong to the same minihalo, we compare the selected dark matter particle IDs associated with each star-formation event. We associate a new star-forming group with a previously identified minihalo if at least $20\%$ of the selected particle IDs overlap. We adopt this threshold to account for changes in the central dark matter particle population as the minihalo evolves, while maintaining sufficient overlap to identify the same structure. Otherwise, the group is assigned to a new minihalo. By assigning each sink particle to a minihalo at the time of its formation, we can distinguish stars that initially form in separate dark matter structures, even if their parent minihalos subsequently merge.

We estimate the masses of these minihalos using the spherical-overdensity definition $M_{200\mathrm{m}}$, where $R_{200\mathrm{m}}$ is the radius within which the mean enclosed density is $200$ times the mean matter density of the Universe at that redshift,

\begin{equation}
\frac{M_{200\mathrm{m}}}{(4\pi/3)R_{200\mathrm{m}}^3}
=200\,\bar{\rho}_{\mathrm{m}}(z).
\end{equation}

We calculate the enclosed dark matter mass directly from the simulation particles and estimate the total halo mass using the cosmic baryon fraction, $M_{200\mathrm{m}}\approx M_{\rm DM}(<R_{200\mathrm{m}})\,\Omega_{\rm m}/(\Omega_{\rm m}-\Omega_{\rm b})$.

Star formation does not occur within the same single dark matter minihalo throughout the simulation. Instead, new groups of stars form in several distinct dark matter structures with estimated masses of $M_{200\mathrm{m}}\sim10^6$--$10^7\;M_{\odot}$. We identify five distinct star-forming groups, each associated with a different minihalo at the time of formation. We number these minihalos chronologically according to when they first form stars, beginning with Minihalo 1 at $z\approx16.03$ and ending with Minihalo 5 at $z\approx7.10$. Minihalos 1 and 3 subsequently merge during the course of our simulation, bringing initially separate stellar populations into the same dark matter halo.

\subsection{Binary Merging} \label{subsec:binary merging}

To allow the simulation to run for longer times, we artificially merge close binary systems during their evolution, to prevent the simulation time-stepping from becoming arbitrarily small. We merge two sink particles if they have an orbital separation below the adopted merger radius, which is set by the sink gravitational softening scale and a minimum binary orbital period of $100$ years (corresponding to Keplerian semi-major axes of approximately $30-60$ au for equal-mass binaries with individual stellar masses of $1-10 \; \rm M_{\odot}$), are gravitationally bound (binding energy $E_{\rm bind} < 0$), have component masses within an order of magnitude of one another ($q \geq 0.1$), have reached the main sequence, and have an accretion timescale $M/\dot{M} \geq 2t_\star$, where $M$ is the sink mass, $\dot{M}$ is the sink accretion rate, and $t_\star$ is the stellar age. When these conditions are satisfied, we replace them with a single sink particle. We set the merged sink mass and momentum equal to the sum of their components, while the metallicity and age are calculated as the mass-weighted averages of the progenitors' values.

This merging prescription reduces the computational cost associated with very tight binaries, which would otherwise require extremely small time-steps. The characteristic sink-particle timestep can be as short as $\sim 10^{-3}$ yr before close binaries are merged, increasing to $\sim 10^2$ yr after merging. This numerical limitation is not unique to our simulation. Previous Pop.~III simulations have generally followed protostellar systems for much shorter timescales ($\sim10^3-10^5$ yr) than the $\sim500$ Myr evolution followed here \citep[e.g.,][]{susa_mass_2014, stacy_building_2016, hirano_formation_2017, latif_birth_2022, sharda_magnetic_2025}. Resolving tight multiple systems over substantially longer timescales becomes computationally expensive because their short orbital periods require correspondingly small time-steps. Our merging prescription provides a practical way to address this numerical limitation without imposing a fixed minimum timestep, allowing us to follow the stellar population over much longer timescales.

The merging performed in this simulation is a numerical treatment rather than a physical model of stellar mergers. The stars that we merge form nearly synchronously in time and have extremely short orbital periods compared to their protostellar evolution timescales. As a result, the mass-weighted average age provides a reasonable approximation for the age of the combined system. However, some systems that are artificially merged in the simulation could instead remain as close binaries in reality.

The merger criteria are designed to minimize the effect of this approximation on the stellar properties that we measure. We do not merge (proto)stars until they have reached the main sequence, allowing their masses and ages to be followed individually through most of their (proto)stellar evolution. We also note that massive stars can continue accreting after reaching the main sequence. For this reason, we additionally require the accretion timescale to satisfy \(M/\dot{M} \geq 2t_\star\) before allowing a merger. This prevents stars that are still rapidly accreting from being merged and restricts mergers to systems whose masses are evolving more slowly. The binary properties presented later in this paper are also measured while the stellar components remain individually resolved, before an artificial merger occurs. 

Artificial merging can nevertheless affect the subsequent evolution of these systems. If the components remained separate, their later orbital evolution and dynamical interactions could differ, potentially changing the surviving binary population or producing additional stellar ejections. Their subsequent accretion and feedback histories could also differ once the individual stars are replaced by a single sink. We therefore expect measurements made before artificial merging to be less sensitive to this prescription than the later dynamical and feedback evolution of the system.

One concern with artificially merging close binaries is that the subsequent supernova feedback may differ from that of the individual progenitor stars. In particular, combining multiple progenitors into a single sink can change the timing and total ejecta energy associated with later supernova events. However, the characteristic energy scale of a core-collapse supernova adopted in our simulation is of order \(10^{51}\,\mathrm{erg}\), and a system with multiple massive progenitors would contribute approximately this energy per progenitor. For example, a system with two progenitors would correspond to a total supernova energy of order \(2\times10^{51}\,\mathrm{erg}\), which is within the range of energies predicted for massive, low-metallicity, and Pop.~III supernovae \citep{nomoto_population_2004, nomoto_nucleosynthesis_2006, heger_nucleosynthesis_2010, chiti_enrichment_2026}. We therefore do not expect the artificial merging to introduce an order-of-magnitude error in the total supernova energy budget, although the detailed timing, ejecta masses, and energy distribution among individual events may be affected. The explosion energies of Pop.~III supernovae remain uncertain, with theoretical and observationally motivated estimates spanning approximately \(\sim10^{51}\)--\(10^{53}\,\mathrm{erg}\). Given this existing uncertainty, the change in the total ejecta energy introduced by combining a small number of progenitors is unlikely to dominate the overall uncertainty in the supernova feedback energetics.


The code also includes an additional merging mechanism at sink formation: if two sinks are extremely low mass (of order a few particle masses), form nearly simultaneously, and lie within each other's interaction kernel, they may be merged immediately to suppress spurious sink formation. These events may correspond to systems that would physically form very tight binaries, with separations of order a few AU (set by the sink gravitational softening kernel radius), but are unresolved at the simulation resolution and therefore treated as a single sink from their formation.

As a result, sinks with \(N_{\rm prog} > 1\), where $N_{\rm prog}$ is the number of progenitor sink particles that merged to form the current sink, can reflect either unresolved early mergers or sub-timestep merging events, rather than the late-time evolution of well-resolved binaries. We account for this by tracking progenitor counts and incorporating them into the IMF and multiplicity analysis, so that such systems are not artificially excluded. As in any finite-resolution simulation, very close binaries near or below the gas cell resolution limit may remain unresolved. The exact effect of these unresolved systems is difficult to predict because their subsequent evolution and mergers are nonlinear. However, at the resolution of our Pop. III simulation, missing such close binaries is not expected to produce an order-unity correction to the IMF or binary fraction. The effect may be larger near the IMF turnover, where the results are more sensitive to resolution.

\section{\label{sec:results} Results}
\subsection{Gas Evolution and Feedback}

Figure \ref{fig:visualizations} shows the evolution of the gas environment over the course of the simulation. Star formation occurs in dense gas filaments and molecular clouds within the cosmological halo. The visualizations capture a large dynamic range and multiple physical processes, including protostellar jets and supernovae.


Protostellar jets launch outflows that evacuate cavities in the surrounding gas before the stars reach the main sequence, visible in the kinetic energy projections in Figure \ref{fig:visualizations}. In our simulation, stars within the adopted supernova mass range ($1.4$--$20\,M_\odot$) undergo supernova explosions at the end of their main-sequence lifetimes, producing remnants that sweep up the surrounding interstellar medium, as visible in the later stages of Figure \ref{fig:visualizations}. Overall, the simulation evolves for $\sim 500$ Myr.

Figure~\ref{fig:chemistry visualizations} (top three panels and bottom-left panel) shows the column-averaged abundances of H$_2$, HI, HII, and the electron fraction $x_e$ in a $20$ pc region centered on a group of sink particles at one representative snapshot ($z = 16.02$) during the $\sim 500$ Myr evolution. The region contains $16$ sink particles, including six main-sequence stars (stage 5) and ten protostars at earlier evolutionary stages (stages 1, 2, and 4). The gas is predominantly neutral, with atomic hydrogen fractions of $x_{\rm HI} \sim 0.93\text{--}1$ over most of the region. A localized HII region surrounds the sink particles, where both the HII abundance and electron fraction are enhanced. The ionization is likely driven primarily by stellar ionizing radiation, particularly from the main-sequence stars, while protostellar jets may also influence the distribution of the surrounding gas. In contrast, the H$_2$ map traces dense molecular gas concentrated in clumps around the ionized region. 

Figure~\ref{fig:chemistry visualizations} shows HII visualizations (bottom-middle and bottom-right panels) across different regions of the simulation on progressively smaller scales around the (proto)star. These maps show the ionized gas structure on scales of $0.1$ pc and $0.01$ pc, illustrating how the HII structure changes closer to the central sink particle.

\subsection{\label{subsec:stellar pop overview} Formation History of the Stellar Population}

\begin{figure}
\centering
\includegraphics[width=\columnwidth]{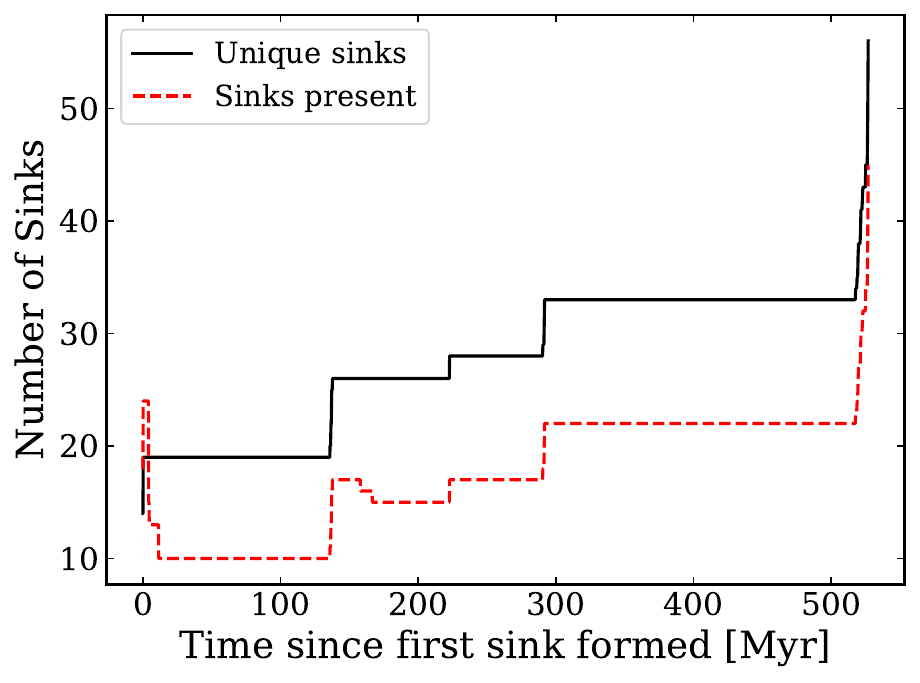}
\caption{Evolution of the sink population as a function of time since the formation of the first sink particle. We plot both the cumulative number of unique sinks formed and the total number of progenitors represented by sink particles present at each snapshot. The dashed red line is calculated by summing the progenitor counts of the sink particles in each snapshot, rather than counting the sink particles themselves. The progenitor count accounts for merger events that are not individually resolved, including numerical mergers and physically close systems that merge below the resolution limit (see Section \ref{subsec:binary merging}). Because some progenitors were never identified as separate sink particles in the stored snapshots, the red curve can exceed the cumulative number of unique sinks. The step-like behavior reflects discrete sink formation events over time.}
\label{fig:sink evolution}
\end{figure}

We track the number of unique sink particles, their masses, and their formation histories. In total, $56$ unique stars form over the course of the simulation, including those that are later artificially merged. This count excludes additional unresolved progenitors associated with mergers near sink formation, which are discussed in Section \ref{subsec:binary merging}. In Figure \ref{fig:sink evolution}, we also show the total number of progenitors represented by the sink particles present at each snapshot (dashed red line). The black curve excludes additional unresolved progenitors associated with mergers at sink formation, while the red curve includes these progenitors through the progenitor counts of the surviving sink particles.

The progenitor count accounts for systems that are artificially merged according to the binary merger criteria described in Section \ref{subsec:binary merging}. In addition, the code allows sink--sink mergers under specific conditions: both sinks are extremely low mass (of order a few particle masses), have just formed, and are located within an interaction kernel of each other at nearly the same formation time. This is implemented to suppress spurious sink formation in cases where multiple density peaks occur within a small spatial region and cannot be numerically distinguished as separate objects. While such cases may physically correspond to close binaries, they are effectively unresolved from the moment of formation and are therefore treated as a single sink.


Figure \ref{fig:sink evolution} shows that star formation occurs in discrete episodes rather than continuously. Within these episodes, several sinks often form within a short interval of one another. Periods of rapid sink formation are separated by intervals with little or no new sinks. This is consistent with previous Pop.~III simulations, which find that star formation happens in bursts due to fragmentation and time-variable accretion \citep{prole_fragmentation-induced_2022}. 


As seen in Figure \ref{fig:sink evolution}, the number of progenitors represented by surviving sink particles at a given time does not increase monotonically. Instead, it fluctuates, which can reflect the removal of sink particles through processes such as supernovae. Artificial merging (see Section \ref{subsec:binary merging}) does not necessarily reduce this count, since the surviving sink can retain the progenitor counts of its components. At late times, the progenitor count in surviving sinks (red curve) is lower than the total number of unique sink particles formed (black curve).

In Figure \ref{fig:sink evolution}, we see that around half of the stars form within $400$ Myr of the first sink forming. A sharp increase in sink formation then occurs approximately $500$ Myr after the formation of the first sink. These sinks form in different regions of the ultra-faint dwarf halo (see Figure \ref{fig:visualizations}; bottom right). This later increase in star formation is consistent with hydrogen and helium gas being accreted from the intergalactic medium or redistributed within the halo, allowing gas in these new regions to cool and reach star-forming densities. Importantly, this later-forming gas remains predominantly metal-free and does not appear to be significantly enriched by the earlier supernovae. The stars that form in these regions therefore remain Pop.~III stars.

We also check the metallicities of the 56 stars at the snapshots used to measure their stellar masses. Of these, 54 remain strictly metal-free ($Z=0$), while two have trace metallicities of $Z \simeq 2.6\times10^{-7}$ ($Z/Z_\odot \simeq 1.8\times10^{-5}$). Thus, the stellar population remains overwhelmingly metal-free, with only two stars showing very small levels of enrichment. These two stars are already trace-enriched at their first appearance in the simulation, suggesting that they formed from gas with trace levels of enrichment. Both of the trace-enriched stars have very similar masses of $\sim 3.8 \; M_{\odot}$. 

While Pop.~III stars are conventionally defined as forming from metal-free gas \citep{bromm_first_2004,klessen_first_2023}, the transition between Pop.~III and Pop.~II star formation need not necessarily correspond to a single well-defined metallicity threshold. Instead, the critical metallicity depends on the relevant cooling and fragmentation mechanisms. Gas-phase metal-line cooling gives a commonly discussed critical metallicity of $Z_{\rm crit}/Z_\odot\sim10^{-3.5}$, while dust-induced fragmentation can become important at substantially lower metallicities, around \(Z_{\rm crit}/Z_\odot\sim10^{-5.5}\) \citep{smith_why_2024}. Consistent with this, \citet{chon_transition_2021} find that the number of low-mass stars increases for \(Z/Z_\odot\gtrsim10^{-5}\), with dust-induced fragmentation producing stars with masses of $\sim0.01-0.1\;M_\odot$, although the overall stellar mass distribution remains top-heavy until considerably higher metallicities. The two trace-enriched stars in our simulation, with \(Z/Z_\odot\simeq1.8\times10^{-5}\), therefore lie in the regime between these commonly discussed dust- and gas-phase critical metallicities. We therefore refer to these two objects as trace-enriched rather than assigning them a strict Pop.~II classification. We retain both stars in the IMF and other population-level analyses presented here.

We also quantify the star formation efficiency (SFE) of the first star-forming minihalo. For the first halo that forms stars, the total stellar mass that forms is $M_{\star} \approx 320 \; \rm M_{\odot}$. Using the baryonic mass estimated from the halo's dark matter mass and the cosmic baryon fraction at the onset of star formation, $M_{b} \approx 4.7\times 10^{5} \; \rm M_{\odot}$, we estimate the SFE as $\epsilon_{\rm SFE} = \frac{\rm M_{\star}}{\rm M_{b}} \approx 6.8\times10^{-4}$. Observational constraints on Local Group dwarf galaxies imply early SFEs of $\epsilon_{\rm SFE}\sim 3\times10^{-3}$--$5\times10^{-3}$ for halos of $M_{\rm dm}\sim10^{7}$--$10^{9}\;\rm M_{\odot}$ at $z\sim6$ \citep{kang_ghostly_2019}, several times higher than the efficiency measured for our first star-forming halo. The low efficiency in our simulation indicates that only a small fraction of the available gas is able to form stars before feedback and other processes suppress further star formation.




The absence of very massive stars ($> 200 \; \rm M_{\odot}$) in our simulation, in contrast to other simulations \citep[e.g.,][]{mckee_formation_2008, hosokawa_protostellar_2011, hirano_one_2014, susa_mass_2014}, likely reflects differences in the included physics. As discussed in \citet{meziani_forged_2026}, the simulation includes a broad range of processes that regulate stellar masses, including infrared (IR) radiative cooling, H$_2$ cooling, protostellar winds, radiative feedback from massive stars (including UV photoionization and photoheating), and core-collapse supernova feedback. These processes limit accretion and remove gas from star-forming regions, which may contribute to the absence of stars more massive than $200 \; \rm M_{\odot}$. 

On the other hand, we do see approximately nine stars with masses near or below a solar mass. A useful question is whether any of these low-mass stars could survive to the present day. A star that formed at $z\sim16$ would need to survive for $13.55$ Gyr to live to the present day. 
Using the approximate main-sequence scaling $t_{\rm MS}\sim M/L$, together with the mass-luminosity relation $L\propto M^{3.5}$, gives
\begin{equation}\label{eq:main-sequence lifetime}
t_{\rm MS} \approx 10 \; \mathrm{Gyr}\left(\frac{M}{M_\odot}\right)^{-2.5}
\end{equation}
\citep{eddington_relation_1924}. Solving Equation~\ref{eq:main-sequence lifetime} with a main-sequence lifetime of $13.55$ Gyr gives \(\frac{M}{M_\odot} \approx \left(\frac{10}{13.55}\right)^{0.4} \approx 0.89.\) This estimate assumes that low-mass Pop. III stars approximately follow the standard main-sequence mass–luminosity relation; detailed zero-metallicity stellar-evolution models may shift the precise survival threshold \citep{ferreira_evolution_2026, ferreira_red-giant_2026}.
Thus, under this approximation, only stars with masses below roughly $0.9 \; \rm M_{\odot}$ would be expected to survive from $z \sim 16$ to the present day. This is broadly consistent with previous studies that consider Pop.~III stars with masses $\lesssim 0.8 \; \rm M_{\odot}$ as potential present-day survivors \citep{hartwig_constraining_2015, magg_observational_2019}. In our sample, $8$ stars fall below this approximate mass threshold. However, these stars lie below our estimated completeness limit of $\sim 1 \; \rm M_{\odot}$, so the exact number of surviving low-mass stars is not well constrained. Previous STARFORGE resolution studies find that increasing the numerical resolution primarily increases the number of low-mass stars, while leaving the high-mass end of the IMF comparatively unchanged \citep{guszejnov_effects_2022}. Thus, a higher-resolution simulation may produce additional low-mass ($\rm M_{\star} \lesssim 1 \; M_{\odot}$) Pop.~III stars.

\subsection{\label{subsec:binaries} Binary Statistics}

\begin{figure*}
    \centering
    \includegraphics[width=\textwidth]{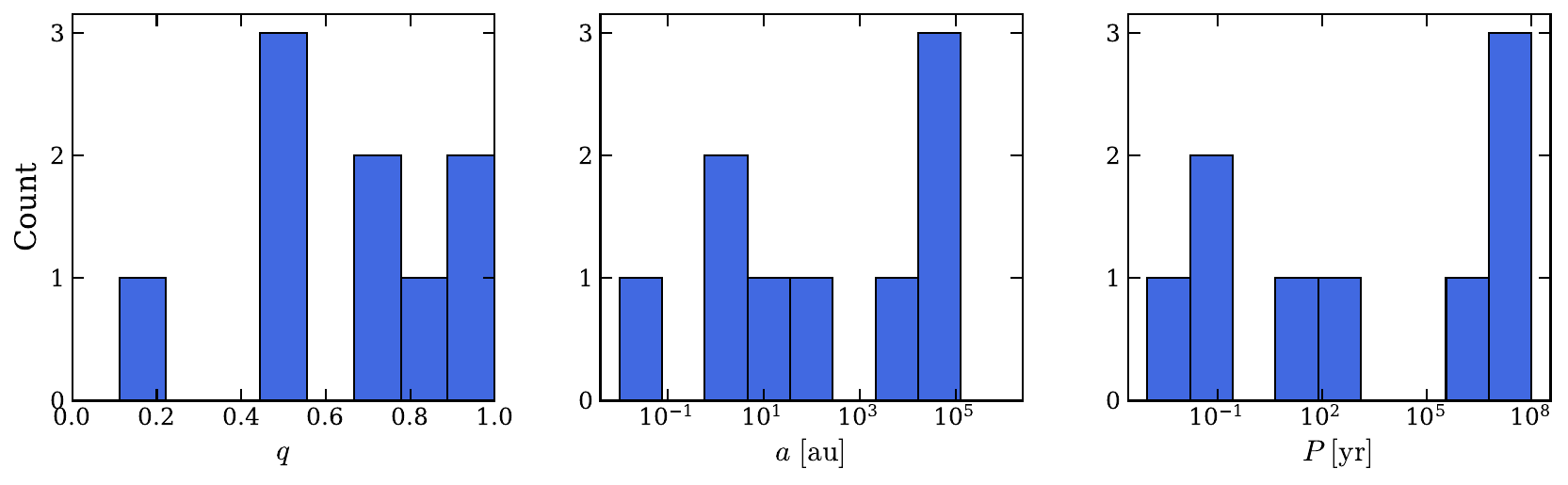}
    \caption{Distribution of orbital properties for the bound pairs identified in binary and higher-order multiple systems in the simulation. The left panel shows the mass ratio, $q$, defined as the ratio of the less massive component to the more massive component at each level of hierarchy. The middle panel shows the semi-major axis in au, and the right panel shows the corresponding orbital periods in years. For higher-order multiple systems, we calculate the orbital properties using the combined mass and center of mass of any inner bound pair.} 
    \label{fig:binary stats}
\end{figure*}

Once a (proto)stellar multiple system has formed, we identify gravitationally bound systems using their orbital energy. We first identify which pairs of (proto)stars are gravitationally bound by evaluating their orbital energy. We do this by looking for systems of (proto)stars which are bound and have negative orbital energy, evaluated in the center-of-mass (COM) frame of the stellar multiple. For every pair of stars, the COM velocity is calculated using a mass-weighted average of the velocities for each pair of stars. The stellar velocities are then transformed into the COM frame by subtracting the COM velocity, which yields the relative velocity.

The kinetic energy in the COM frame is computed from the relative velocity,
\begin{equation}
K = \frac{1}{2} \mu |\textbf{v}'_1 - \textbf{v}'_2|^2,
\end{equation}
where $\mu = \frac{m_{1}m_{2}}{m_{1}+m_{2}}$ is the reduced mass for the pair of stars and $\textbf{v}'_1$ and $\textbf{v}'_2$ are the velocities of the two (proto)stars in the COM frame. The gravitational potential energy is calculated using
\begin{equation}
U = \frac{-G m_1 m_2}{r},
\end{equation}
where $G$ is the gravitational constant, $r$ is the spatial separation between the two (proto)stars, and $m_1$ and $m_2$ are the masses of the two (proto)stars. The total orbital energy of the pair is defined as
\begin{equation}
E_{\rm orb} = K + U,
\end{equation}
and pairs with $E_{\rm orb} < 0$ are classified as gravitationally bound.

Higher-order systems are identified hierarchically. We first calculate the orbital energies of all possible pairs and identify the most-bound pair, corresponding to the pair with the most negative orbital energy. Once two stars form a bound pair, that pair is treated as a single subsystem located at its center of mass, with a total mass equal to the sum of the two stellar masses. After replacing this pair with its COM subsystem, we recalculate the orbital energies between all remaining stars and subsystems and again identify the most-bound pair. If another bound pair is identified, it is similarly replaced by its COM subsystem. This procedure is repeated until no remaining pair has negative orbital energy. For a higher-order system to be identified, the additional star or subsystem must be bound to the COM of the inner bound subsystem.


For bound orbits, we determine the semi-major axis $a$ for each hierarchical pair

\begin{equation}
a = - \frac{Gm_1 m_2}{2 E_{\rm orb}}.
\end{equation}

Using Kepler's third law, the orbital period $P$ can be computed using 

\begin{equation}
P = 2 \pi \sqrt{\frac{a^3}{G(m_1 + m_2)}}.
\end{equation}

The mass ratio for each hierarchical pair is defined as 
\begin{equation}
q = \frac{\min(m_1,m_2)}{\max(m_1,m_2)}.
\end{equation}
For higher-order systems, $m_1$ or $m_2$ may represent the total mass of a bound subsystem rather than the mass of a single star. Similarly, the separation and orbital properties of an outer hierarchical pair are calculated using the COM positions and velocities of the corresponding subsystems. 

Over the course of the simulation, we identify 56 unique stars. We identify 6 binary systems and 1 quadruple system based on the orbital properties of the sink particles. Separately, a single sink particle can already represent more than one progenitor star when it forms because of unresolved early merging. These additional progenitors are contained within a single sink particle and are therefore not treated as separate companions when identifying binary and higher-order multiple systems. Following standard definitions, we define the multiplicity fraction as $MF = \frac{B+T+Q+...}{S+B+T+Q+...}$, and the companion frequency as $CF = \frac{B + 2T + 3Q + ...}{S+B+T+Q+...}$, where $S, B, T \; \rm and \; Q$ are the numbers of single, binary, triple and quadruple systems, respectively \citep{guszejnov_effects_2023, offner_origin_2023}. For our $56$ unique stars, we identify $40$ single systems, $6$ binaries, and $1$ quadruple system, giving $MF = 7/47 \approx 0.15$ and $CF = 9/47 \approx 0.19$. 


We also consider the additional stellar progenitors represented within individual sink particles. As mentioned above, a single sink particle can represent more than one progenitor star at formation because of unresolved early merging. These systems could represent physical binaries. Overall, including all the progenitors, we identify one multiple system with $9$ members, one triple system, and $5$ binary systems. Including these additional progenitors increases the total number of individual stars from $56$ to $62$. Using the same system-based definitions as the paragraph above, we find $MF = 7/47 \approx 0.15$ and $CF = 15/47 \approx 0.32$. 


Our simulation is qualitatively consistent with previous Pop.~III simulations in producing a substantial population of multiple systems, although the inferred multiplicity is somewhat lower than in \citet{stacy_constraining_2013}. STARFORGE simulations of MW-like molecular clouds find that the multiplicity fraction increases with stellar mass, with most high-mass stars belonging to multiple systems while most low-mass stars are single \citep{guszejnov_effects_2023}. This trend is also supported by observations \citep{offner_origin_2023, noauthor_multi-star_2024, li_observations_2024}. In our simulation, our highest-order multiple system contains several of the most massive stars in our sample, with the quadruple consisting of stars with masses of approximately $110, 28, 15$ and $8$ M$_\odot$. Of the nine stars with masses $M_\star \geq 8\,M_\odot$, five belong to the identified multiple systems ($\sim 56\%$). However, the second-most-massive star in the simulation, with $M_\star \approx 64\,M_\odot$, is not among these systems. For comparison, \citet{stacy_constraining_2013} define the binary fraction as $f_{\rm B}=B/(S+B)$ and find $f_{\rm B}=0.36$. They also report that $53\%$ of individual stars have a binary companion, calculated as $2B/(S+2B)$. Using the analogous star-based quantity for our sample, we find that $16/56\simeq29\%$ of stars belong to a binary or higher-order multiple system. Thus, the fraction of stars with companions in our simulation is lower than that found by \citet{stacy_constraining_2013}, although both simulations produce a substantial population of multiple systems.

Figure \ref{fig:binary stats} shows the distribution of orbital properties for all unique multiple systems identified in the simulation. Binary properties are measured at the latest snapshot in which the system remains resolved as separate sink particles prior to any artificial merging (see Section \ref{subsec:binary merging}). For each system, we select the snapshot at which the companion stars reach a combination of their highest protostellar stage and mass while still remaining distinct, and compute the mass ratio, binary separation, and orbital period at that time. For higher-order multiple systems, we decompose each system into the hierarchical pairs identified through the binding-energy procedure described above and compute the orbital properties of each pair independently.

The mass ratio distribution (Figure \ref{fig:binary stats}: left panel) includes both near-equal-mass systems ($q \gtrsim 0.8$) and unequal pairs ($q \lesssim 0.5$). \citet{stacy_constraining_2013} find a distribution that favors unequal-mass (\enquote{low-$q$}) binaries, whereas this simulation shows a preference for near-equal-mass systems. This difference may arise because the binaries in \citet{stacy_constraining_2013} are evolved for only a few thousand years, potentially limiting their long-term mass evolution. However, the number of multiple systems in this simulation is small, so the inferred distribution may not be statistically representative.

The binary semi-major axis distribution peaks at $\sim 10^{2}\,\mathrm{au}$, with a broad range spanning $\sim 10^{-1} \, \rm au$ to $10^{5}\,\mathrm{au}$. This wide range reflects both the initial conditions of fragmentation and subsequent dynamical evolution. Some of the inferred binary semi-major axes fall below
the sink gravitational-softening scale of $\sim 5$--$10\,\rm au$.
This is possible because gravitational softening does not
impose a minimum separation between sink particles, and
some of these systems are measured shortly before artificial
merging \citep{grudic_starforge_2021,guszejnov_effects_2023}.



\subsection{\label{subsec:stellar ejection} Stellar Ejections}
\begin{figure*}[t!]
    \centering
    \includegraphics[width=0.9\textwidth]{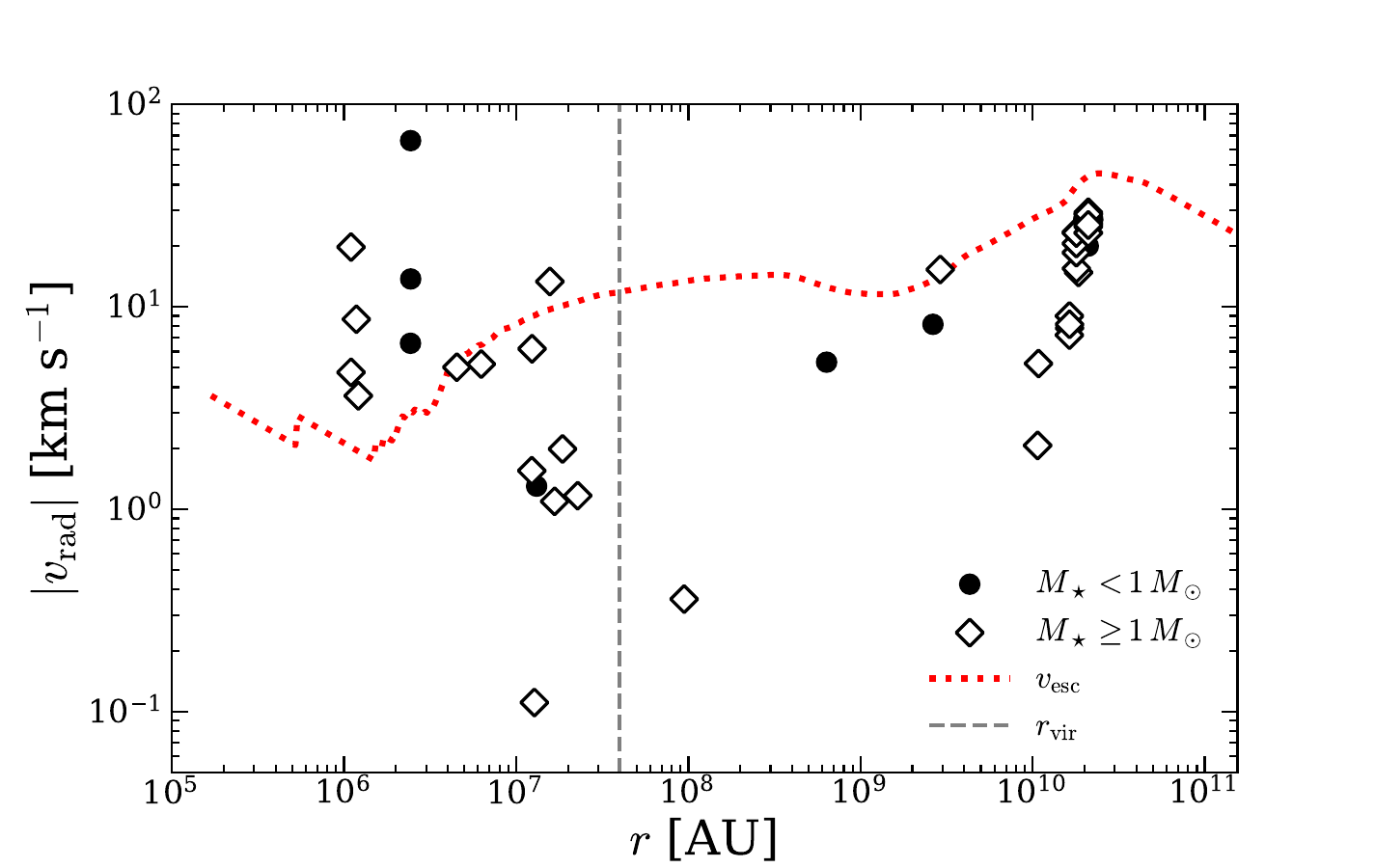}
    \caption{Absolute radial velocity $v_{\rm rad}$ of stars as a function of physical distance from the halo center, at the time they each reach the main sequence or their highest protostellar stage if they do not reach the main sequence. The black points correspond to sink particles with stellar mass $\rm M_{\star} < 1 \; \rm M_{\odot}$, while the diamonds correspond to sink particles with stellar mass $\rm M_{\star} \geq 1 \; \rm M_{\odot}$. The red dotted line shows the reference escape-velocity profile, $v_{\rm esc}$, calculated at $z\approx 7.04$. The vertical gray dashed line shows the approximate virial radius of the first star-forming minihalo.}
    \label{fig:ejections}
\end{figure*}


Close (proto)stellar interactions can lead to ejections from binary or higher-order multiple systems. Following the analysis of \citet{stacy_building_2016}, we examine which (proto)stars remain gravitationally bound to the first star-forming halo as well as the overall global potential. We identify the first star-forming minihalo and track its center throughout the simulation using a fixed set of dark matter particles from its central core. At each snapshot, we update the estimated minihalo center based on the position of the local dark matter density maximum. For each (proto)star, we calculate its position and velocity relative to the minihalo at the snapshot when it reaches the main sequence, or at the highest protostellar stage it reaches if it does not reach the main sequence. To provide a common reference potential for the full stellar population, we calculate a single escape-velocity profile using the latest snapshot represented in this sample, at $z\approx 7.04$. We calculate the enclosed mass as a function of radius from the minihalo center, $\rm M_{\rm enc}(<r) = M_{DM}(<r) + M_{gas}(<r)$, and estimate the escape velocity as

\begin{equation}\label{eq:escape}
v_{\rm esc}(r) = \sqrt{\frac{2G\rm M_{enc}(<r)}{r}},
\end{equation}

where $G$ is the gravitational constant and $r$ is the distance from the halo center. For each star, we evaluate this reference profile at its measured distance from the first star-forming halo. 

We classify a (proto)star as unbound when its velocity relative to the minihalo, $v_{\rm rel} = |v_{\star} - v_{\rm minihalo}|$, exceeds $v_{\rm esc} (r)$ ($v_{\rm rel} > v_{\rm esc}(r)$). Figure \ref{fig:ejections} shows the absolute radial velocity, $|v_{\rm rad}|$, as a function of distance from the halo center.


As shown in Figure \ref{fig:ejections}, (proto)stars with $M_{\star} < 1 \; M_{\odot}$ are shown as circles, while (proto)stars with $M_{\star} > 1 \; M_{\odot}$ are shown as diamonds. The red dotted line is the escape velocity profile, $v_{\rm esc}(r)$, calculated using Equation \ref{eq:escape}. The vertical dashed line represents the approximate virial radius, $r_{\rm vir}$, of the first star-forming minihalo in the simulation. Stars at radii greater than $\sim 4 \times 10^{7} \; \rm AU$ formed in different minihalos, as discussed further in Section \ref{subsec:feedback reg of final masses}. 

Because the first star-forming minihalo is the most massive minihalo for most of the simulation, we use it as the reference system for this analysis. The escape velocity profile plotted in Figure \ref{fig:ejections} extends beyond the virial radius of this minihalo to larger radii in the simulation. At these radii, the enclosed mass includes material outside the first minihalo, including the surrounding dark matter, gas, and other structures. This allows us to examine whether stars that are unbound from the first minihalo may still remain bound to the larger-scale gravitational environment.

While we formally classify a (proto)star as unbound using $v_{\rm rel} > v_{\rm esc}(r)$, Figure \ref{fig:ejections} also provides a visual indication of which stars are unbound. Stars with $|v_{\rm rad}| > v_{\rm esc}$ are unbound, while stars with $|v_{\rm rad}| \lesssim v_{\rm esc}$ may still be unbound if their tangential velocity is sufficiently large. We find that $9$ of the $56$ (proto)stars have $|v_{\rm rad}| > v_{\rm esc}$. Using the full three-dimensional velocity criterion, $v_{\rm rel} > v_{\rm esc}(r)$, $16$ of the $56$ (proto)stars are unbound relative to the reference escape-velocity profile shown in Figure \ref{fig:ejections}, corresponding to an unbound fraction of $\sim 29 \%$.


The unbound stars have relatively modest velocities, so they may not travel large distances from the system. A star traveling at \(\sim10\;{\rm km\,s^{-1}}\) for \(10\) Gyr would travel \(\sim100\) physical kpc if its velocity remained constant. These stars could therefore still be accreted into a MW-mass halo if the low-mass galaxy progenitor halo/region we simulate was later accreted as a MW satellite. Alternatively, individual stars that escape their parent minihalo could still be accreted into the MW even if the parent halo itself is not, since their relatively modest velocities may allow them to remain gravitationally bound to the surrounding matter. Figure \ref{fig:ejections} also shows that the escape velocity remains significant beyond the virial radius of the first minihalo because of the mass in the surrounding environment. Thus, a star can escape the first minihalo without necessarily escaping the larger gravitational system.

\subsection{\label{subsec:initial mass function} Initial Mass Function}

\begin{figure}[!t]
    \centering
    \includegraphics[width=0.90\columnwidth]{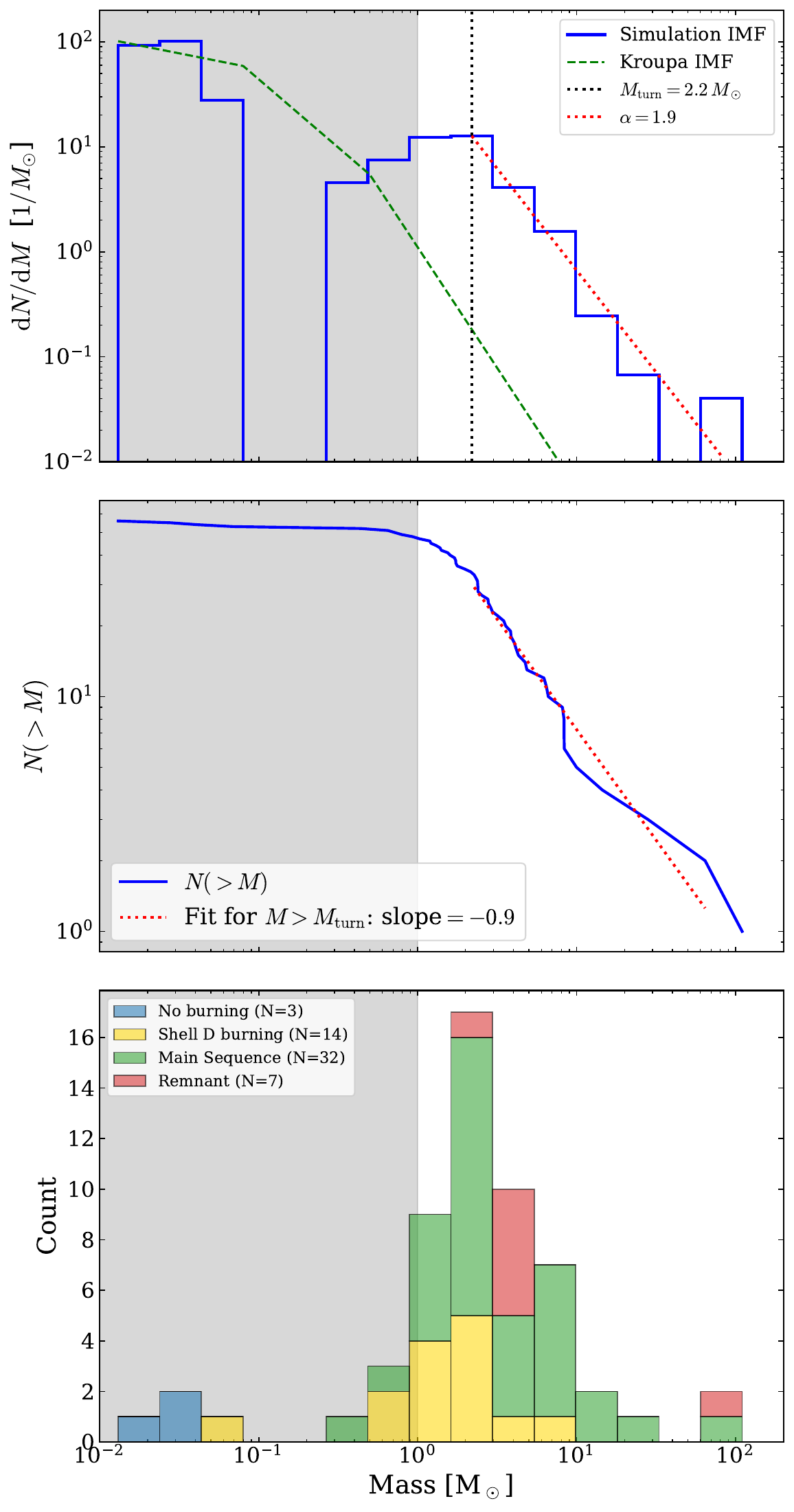}
    \caption{\textit{Top Panel}: IMF of the full population of (proto)stars formed in the simulation, shown as the differential mass distribution $dN/dM$ (blue solid line). We use logarithmically spaced mass bins but divide the number of stars in each bin by its linear mass width, $\Delta M$, to obtain $dN/dM$. The IMF includes all unique (proto)stars formed throughout the simulation, from $z \sim 16$ to $z\sim 7$, using their ZAMS masses or the maximum masses reached within their highest protostellar stages. The dashed green line shows the Kroupa IMF, while the dotted red line shows a single power-law fit. For $dN/dM \propto M_\star^{-\alpha}$, we find $\alpha = 1.9$ for $\rm M_{\star} > 2.2 \; M_{\odot}$.
    \textit{Middle Panel}: Cumulative IMF, shown as the number of (proto)stars with masses greater than $M$, $N(>M)$. The dotted red line shows a power-law fit, and the gray shaded region marks masses below the simulation resolution limit.
    \textit{Bottom Panel}: Distribution of stellar masses used in the IMF for all unique (proto)stars, colored by protostellar stage following \citet{offner_effects_2009}, with remnants defined as described in Section \ref{sec:simulations}. The distribution peaks at $2.4 \; \rm M_{\odot}$, with low-mass objects mostly associated with (proto)stars that have not reached the main sequence.}
    \label{fig:IMF cumulative}
\end{figure}

\begin{figure}[t]
    \centering
    \includegraphics[width=\columnwidth]{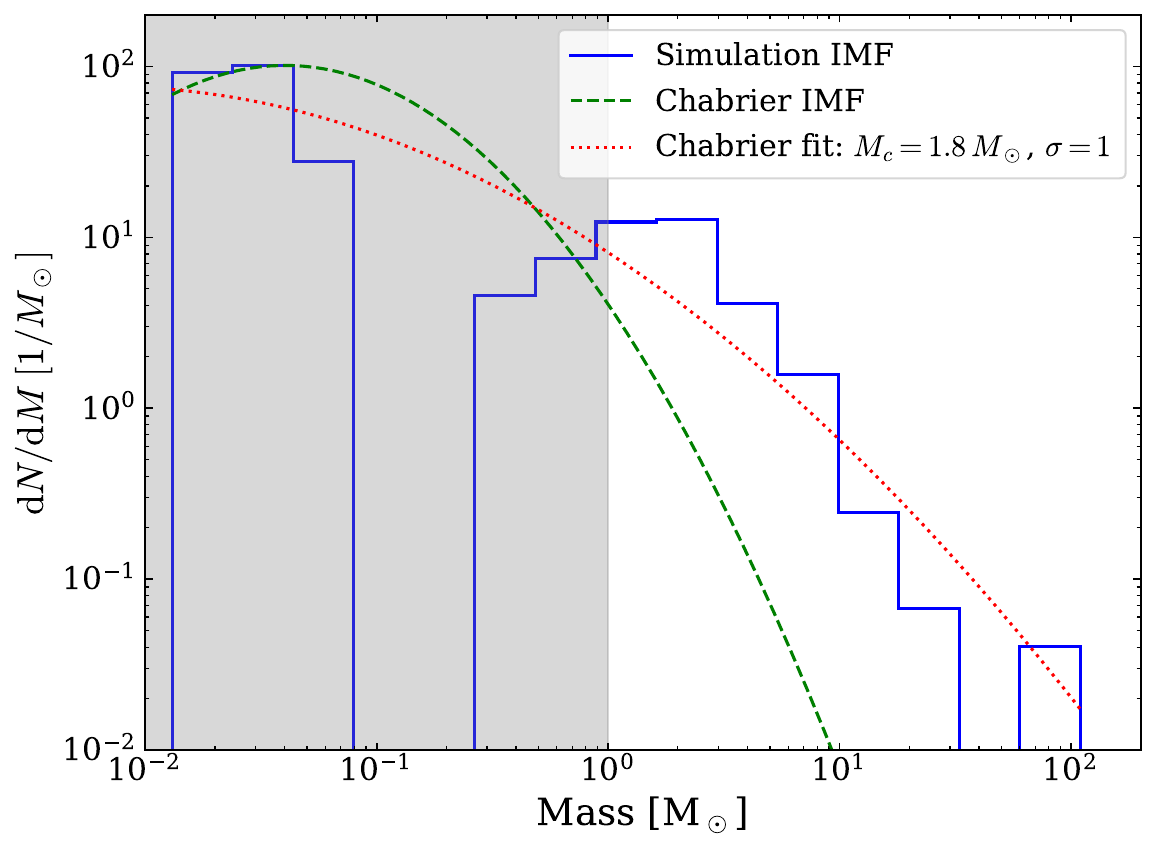}
    \caption{The dashed green line shows the Chabrier IMF, while the dotted red line indicates a best-fit lognormal (Chabrier-type) fit to the simulated IMF. The IMF includes all unique (proto)stars, using their ZAMS masses or maximum masses within their highest protostellar stages.}
    \label{fig:IMF chabrier}
\end{figure}

\begin{figure*}[t]
    \centering
    \includegraphics[width=\textwidth]{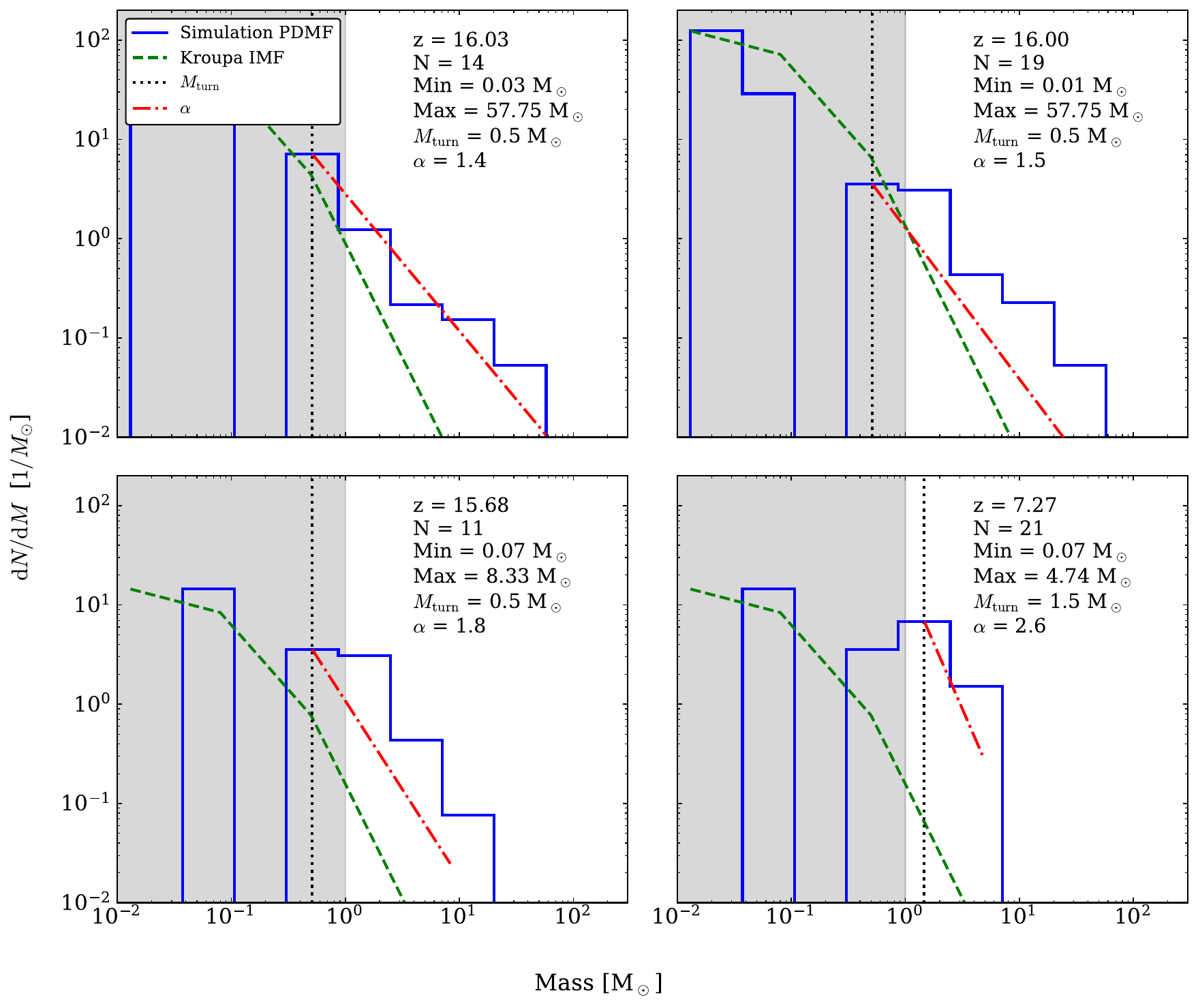}
    \caption{Similar to Figure \ref{fig:IMF cumulative}, this figure shows the time evolution of the present-day mass function throughout the simulation. Each panel shows the sink particle mass distribution at a different simulation output, where $N$ denotes the number of sink particles present at that output. The maximum sink particle mass decreases from $\sim 58 \; \rm M_{\odot}$ at early times of $z\sim16$ to $\sim 5 \; \rm M_{\odot}$ at later times of $z\sim7$, indicating that the later sink population is dominated by lower-mass objects. The turnover mass also increases from $M_{\rm turn}\sim 0.5\;M_{\odot}$ at $z\sim16$ to $M_{\rm turn}\sim 1.5\;M_{\odot}$ at $z\sim7$. }
    \label{fig:IMF over time}
\end{figure*}

\begin{figure*}[t]
    \centering
    \includegraphics[width=0.8\textwidth]{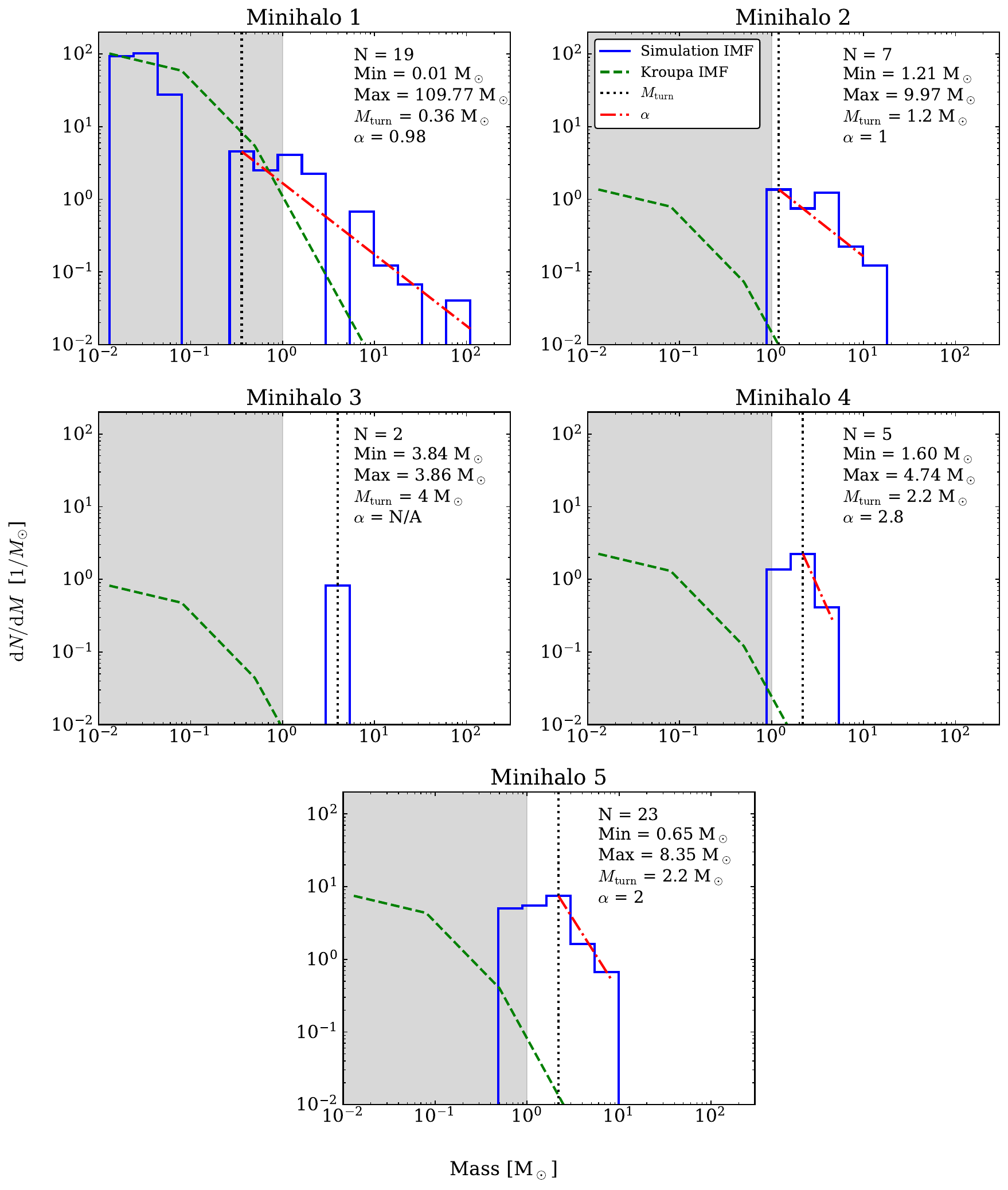}
    \caption{Similar to Figure \ref{fig:IMF over time}, this figure shows the stellar mass distributions, but separates the stellar populations by the halos in which they form. Each panel corresponds to a unique halo and shows the IMF of its sink-particle population. These distributions include all (proto)stars formed in each minihalo throughout the simulation, rather than at a single redshift. We use the ZAMS mass for stars that reach the main sequence, or the maximum mass attained within the highest protostellar stage for those that do not. We use the same notation, symbols, and color scheme as in Figure \ref{fig:IMF over time}.}
    \label{fig:IMF for each different halo}
\end{figure*}

The IMF of the stellar population formed in the simulation is shown in Figure \ref{fig:IMF cumulative}. The IMF is calculated using all the (proto)stars that formed in the simulation. If a star reaches its zero-age main sequence mass (ZAMS mass), we use that mass in the IMF. We emphasize that this is the stellar mass at the time the star reaches the main sequence and does not necessarily represent its final mass, since the star may continue to accrete after reaching the main sequence. If a (proto)star has not reached the main sequence, we instead use the maximum mass attained within the highest protostellar stage it reaches. The bottom panel of Figure \ref{fig:IMF cumulative} shows the distribution of stellar masses used in the IMF for all unique (proto)stars in the simulation. The colors indicate the protostellar stage following the protostellar evolution model of \citet{offner_effects_2009}. The gray shaded region corresponds to the (proto)stars that are not reliably resolved due to the resolution limit of the simulation.

We parameterize the IMF as $dN/dM \propto \rm M^{- \alpha}$ and find a best-fit slope of $\alpha = 1.9$ over the mass range $\rm M_{\star} > 2.2 \; M_{\odot}$. We obtain this slope from a least-squares fit to the cumulative mass distribution, $N(>M) \propto M^{(1-\alpha)}$. For comparison, the Kroupa IMF is described by a broken power law with $\alpha = 0.3$ for stellar masses between $0.01$ and $0.08 \; M_{\odot}$, $\alpha = 1.3$ for masses between $0.08$ and $0.5 \; M_{\odot}$, and $\alpha = 2.3$ for masses greater than or equal to $0.5 \; M_{\odot}$ \citep{kroupa_variation_2001}. Our fitted high-mass IMF slope is only slightly shallower (\enquote{top-heavy}) than that of the Kroupa IMF. From the top panel of Figure \ref{fig:IMF cumulative}, we also see a gap between approximately $0.1$ and $0.3\;M_\odot$. The stars below this gap form early in the simulation, at the same time as the most massive stars in the simulation form. These low-mass stars form as sinks with only a few gas resolution elements near the resolution limit and are dynamically ejected shortly after formation, stopping further accretion and growth.

As mentioned in Section \ref{sec:simulations}, our maximum gas mass resolution is $m_{\rm gas} \sim 10^{-2}\; M_\odot$. Previous STARFORGE resolution studies find that incompleteness in the stellar mass function becomes important below approximately $M_{\rm complete}\sim100\,m_{\rm gas} $ \citep{guszejnov_effects_2022}. Applying this criterion gives a completeness limit of $M_{\rm complete}\sim1\, \rm M_\odot$ for the simulation analyzed in this paper. Because the IMF becomes incomplete near and below $\sim 1 \; M_\odot$, we do not consider the inferred turnover mass to be well converged and instead focus our quantitative interpretation on the IMF above the resolution-limited regime.

In these convergence tests, increasing the numerical resolution primarily adds stars at the low-mass end of the IMF, while the high-mass portion of the mass function remains comparatively stable \citep{guszejnov_effects_2022}. We note, however, that the convergence tests performed in \citet{guszejnov_effects_2022} were done in idealized star-forming clouds, whereas our simulation follows Pop.~III star formation in a fully cosmological environment.

\citet{guszejnov_effects_2022} also found that stars forming in low-metallicity gas, with $Z/Z_{\odot} = 0.01$, produce an IMF that is shallower than allowed by observational uncertainties, with a high-mass slope of $\alpha \sim 2$ for $\rm M_{\star} > 0.5 \; M_{\odot}$. We find a similarly shallow IMF in metal-free gas in this work, with $\alpha = 1.9$, which is qualitatively consistent with the trend seen in those simulations, despite low statistics. Our IMF has a turnover mass of $M_{\rm turn}\sim2.2 \;M_\odot$, which is higher than that found in \citet{guszejnov_effects_2022}. Although our simulation probes the more extreme $Z=0$ regime, it includes only $56$ unique (proto)stars.

We also compare the simulated IMF to a Chabrier-type log-normal form in Figure \ref{fig:IMF chabrier}. The best-fit distribution has a characteristic mass of $M_c \approx 1.8\,M_\odot$ and a logarithmic width of $\sigma \approx 1$. Because the low-mass IMF is incomplete below $\sim 1\,M_\odot$, we do not interpret the fitted turnover as a converged physical prediction. We also show the time evolution of the stellar mass distribution in Figure \ref{fig:IMF over time} using the present-day mass function (PDMF), which represents the mass distribution of sink particles present at each snapshot, rather than the masses of all stars formed throughout the simulation, and the stellar mass distributions separated by their star-forming minihalos in Figure \ref{fig:IMF for each different halo}. We discuss these trends in more detail in Section \ref{subsec:feedback reg of final masses}.

\begin{figure}
    \centering
    \includegraphics[width=\columnwidth]{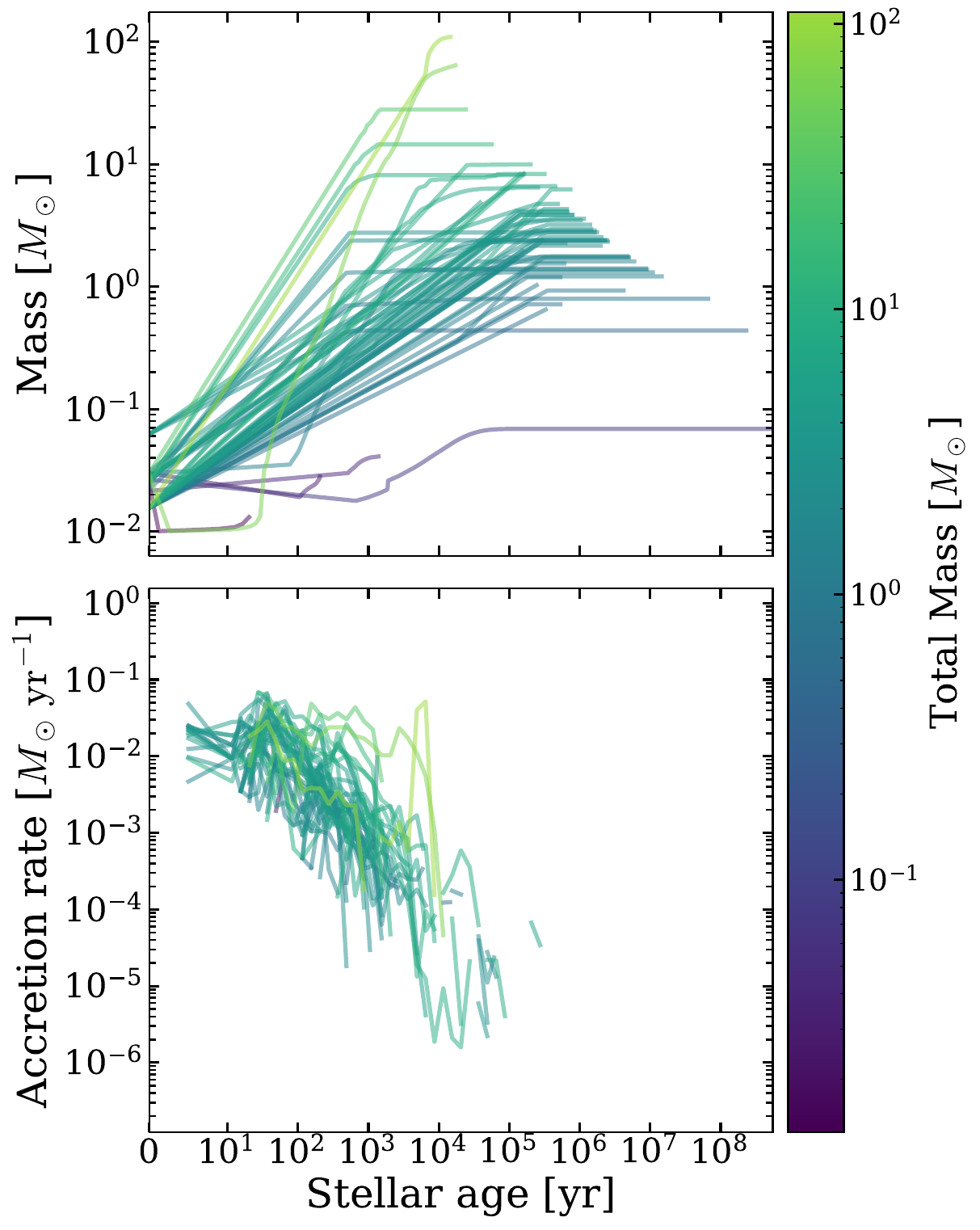}
    \caption{\textit{Top}: Stellar mass as a function of time since formation for each star in the simulation. \textit{Bottom}: Accretion rate as a function of time since formation for each star in the simulation. The x-axis shows the time elapsed since each individual star formed, rather than the global simulation time. The color bar indicates the final stellar mass.}
    \label{fig:mass and mdot vs time}
\end{figure}

\begin{figure*}
    \centering
    \includegraphics[width=\textwidth]{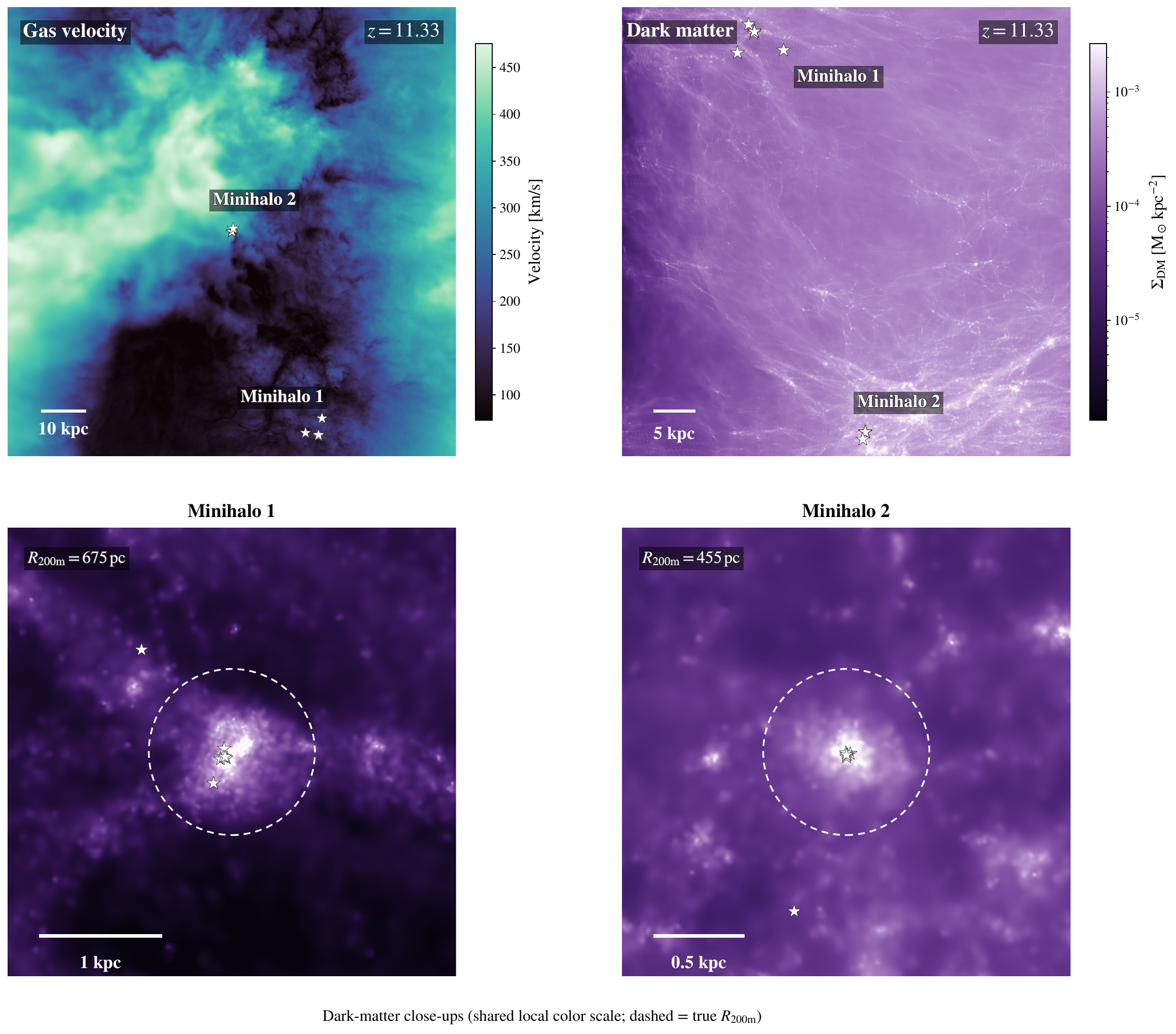}
    \caption{\textit{Top Left}: Projected gas velocity map, with the color bar representing the velocity magnitude. White stars mark the positions of sink particles in the simulation. In chronological order, the first group of stars to form are in Minihalo 1 and the second group of stars to form are in Minihalo 2. 
    This snapshot corresponds to redshift $z = 11.33$. \textit{Top Right}: Projected dark matter surface density, with white stars marking the corresponding sink particle positions. As shown in the figure, the second group of stars forms within a different minihalo. \textit{Bottom Left}: Zoom-in of the dark matter surface density associated with Minihalo 1, using the same color scheme as the top-right panel. The visualization is enlarged for clarity and is therefore not shown to the same spatial or color-bar scale as the top-right panel. The dashed circle represents the approximate size of the dark matter halo, shown to scale. Bottom Right: Same as the bottom-left panel, but showing Minihalo 2 instead.}
    \label{fig:velocity and DM plots}
\end{figure*}


\section{Discussion}\label{sec:discussion}


\subsection{Accretion Histories of Massive Stars} \label{subsec:acc hist of mass stars}

To understand what sets the upper end of the IMF, we examine the growth and accretion histories of the stars in the simulation. Figure \ref{fig:mass and mdot vs time} shows that most of the stars start off having high accretion rates of $\sim 10^{-2} \; \rm M_{\odot}/yr$, which decline by several orders of magnitude during the first million years after their formation. This suggests that the most massive stars experience a phase of rapid early growth, but their final masses are set by when the accretion phase is terminated.


We find that the final mass depends on when and where a sink forms. The most massive stars in the sample tend to form early, when they have more time to accrete and when the local gas reservoir has not yet been strongly affected by fragmentation, stellar dynamics, and feedback. These stars also form in more massive, higher-surface-density gas clouds, which provide larger reservoirs for continued accretion. Later-forming stars therefore reach lower final masses in part because they form in less massive, lower-surface-density clouds, rather than simply because their instantaneous accretion rates are smaller. Within each star-forming region, the available gas can also be redistributed, expelled, or accreted by other sinks. This means that the upper end of the IMF is likely set by both the accretion rate onto massive stars and the duration over which they remain connected to a dense gas supply.

\subsection{Processes Regulating the Initial Mass Function} \label{subsec:feedback reg of final masses}

\begin{table*}[t]
\centering
\caption{Properties of the five star-forming minihalos and their parent gas clouds. The minihalos are numbered chronologically according to when they first form stars. Here, $z_{\rm form}$ is the redshift at which the first sink particle forms in each minihalo. $M_{\rm DM}$ is the dark matter mass enclosed within $R_{200\mathrm{m}}$, where $R_{200\mathrm{m}}$ is the radius within which the mean enclosed matter density is $200$ times the mean matter density of the Universe. Both quantities are evaluated at the onset of star formation in each minihalo. $M_{\rm cloud}$ is the gas mass of the parent star-forming cloud, identified using connected gas with a hydrogen number density of $n_{\rm H} \geq 100\,\mathrm{cm}^{-3}$. The cloud radius $R_{90}$ encloses $90\%$ of the selected cloud gas mass, and $\Sigma_{\rm cloud} = M_{\rm cloud}/(\pi R_{90}^{2})$ is the characteristic cloud surface density. Cloud properties are measured immediately before the first sink forms. Finally, $N_\star$ is the total number of unique stars formed in each minihalo throughout the simulation, while $M_{\star,\max}$ is the maximum stellar mass among those stars, using their ZAMS masses or maximum masses within their highest protostellar stages.}
\label{tab:minihalo_properties}
\small
\setlength{\tabcolsep}{4pt}
\begin{tabular}{lcccccccc}
\hline\hline
Minihalo & $z_{\rm form}$ & $M_{\rm DM}$ & $R_{200\mathrm{m}}$ & $M_{\rm cloud}$ & $R_{90}$ & $\Sigma_{\rm cloud}$ & $N_\star$ & $M_{\star,\max}$ \\
 & & $(10^6\,M_\odot)$ & $(\mathrm{pc})$ & $(10^4\,M_\odot)$ & $(\mathrm{pc})$ & $(M_\odot\,\mathrm{pc}^{-2})$ & & $(M_\odot)$ \\
\hline
1 & 16.03 & 2.36 & 264  & 8.17 & 11.38 & 201.0 & 19 & 109.77 \\
2 & 11.79 & 4.24 & 427  & 1.56 &  6.94 & 103.3 &  7 &   9.97 \\
3 & 10.18 & 17.28 & 780 & 0.61 & 10.06 &  19.1 &  2 &   3.86 \\
4 &  9.23 & 14.65 & 807 & 1.07 &  7.86 &  55.2 &  5 &   4.74 \\
5 &  7.10 & 16.45 & 1059 & 4.16 & 17.37 &  43.9 & 23 &   8.35 \\
\hline\hline
\end{tabular}
\end{table*}

The final stellar masses likely reflect a competition between fragmentation, competitive accretion \citep{zinnecker_prediction_1982, bonnell_competitive_2001, bonnell_origin_2007}, dynamical interactions, and protostellar feedback. Fragmentation increases the number of accreting objects and can divide the available gas reservoir among multiple sinks. Competitive accretion can then allow some stars to grow more rapidly than others, especially those near the center of the potential well. At the same time, protostellar feedback can reduce accretion onto individual stars and limit continued mass growth. 


Figure \ref{fig:IMF over time} plots the PDMF over time and thus helps us see how different physical mechanisms regulate different parts of the IMF. One of the most notable trends is that the PDMF steepens with time. At early times, the PDMF remains relatively shallow above the turnover mass and is top-heavy relative to the present-day \citet{kroupa_variation_2001} IMF, while at later times the mass distribution becomes steeper. Note that in Figure \ref{fig:IMF over time} we plot all the sink particles that we find in the snapshot at that time. 

As discussed in Section \ref{subsec:stellar pop overview}, the stellar population forms in five distinct star-forming groups, each associated with a dark matter minihalo at the time of formation. We number these minihalos chronologically based on when they first form stars, with Minihalo 1 beginning star formation at $z\sim16.03$ and Minihalo 5 being the last to begin forming stars at $z\approx7.10$. Some of these minihalos subsequently merge. Figure \ref{fig:IMF for each different halo} shows the IMF separately for the stars formed in each minihalo. Minihalo 1, which hosts the first group of stars to form, has the broadest stellar mass distribution. Minihalo 1 also forms the most massive star in the simulation ($\rm M_{\star} \approx 110 \; \rm M_{\odot}$). When Minihalo 2 forms, the maximum stellar mass is substantially lower. Minihalo 3 forms only two stars, both during the same star-formation episode. These are also the two trace-enriched stars discussed in Section \ref{subsec:stellar pop overview}, each with $Z/Z_{\odot} \approx 1.8 \times 10^{-5}$, while the stars formed in the other minihalos remain metal-free. The gas in Minihalo 3 was already trace-enriched when these stars formed, potentially due to supernova ejecta from another minihalo. Supernova feedback from these stars may contribute to suppressing further star formation in Minihalo 3. We also see that the IMF slopes in Minihalos 4 and 5 are steeper than in the earlier halos. However, the IMF statistics for the individual minihalos are based on small numbers of stars, so the inferred slopes and differences between minihalos remain somewhat uncertain.

The star-forming minihalos in our simulation have masses of $M_{200\rm m}\sim10^6-10^7\;M_\odot$, below the traditional atomic-cooling threshold ($T_{\rm vir}\sim10^4\;{\rm K}$). Star formation in these halos therefore depends on molecular cooling, particularly through H$_2$. Our simulation includes a redshift-dependent cosmological UV/X-ray background following \citet{faucher-giguere_cosmic_2020}, which adopts hydrogen reionization at $z\sim7.8$. The ionizing background is therefore expected to have a weaker effect during the earliest episodes of star formation at $z\sim16$. However, we do not include a separate cosmological Lyman-Werner (LW) background, which can dissociate H$_2$ and suppress molecular cooling, although self-shielding in dense gas can reduce this effect \citep{chen_radiative_2026}. Such a background could delay star formation until minihalos reach higher masses, potentially allowing larger clouds to accumulate before collapse and producing more massive stars and a more top-heavy IMF. However, the effect on the IMF remains uncertain, as stellar masses also depend on cloud properties, fragmentation, and feedback. 

As discussed in Section \ref{sec:simulations}, our simulation follows the isolated ultra-faint dwarf galaxy \textbf{m09} from the FIRE cosmological simulation suite. The absence of nearby massive galaxies may reduce the contribution of external LW radiation from local sources, resulting in a weaker local LW radiation field. However, more distant galaxies could still contribute to a diffuse cosmological LW background. Given the isolated environment of our simulation, we consider neglecting an additional external LW background a reasonable approximation.

To investigate whether the maximum stellar mass in each star-forming group is related to the properties of its parent GMC, we estimate the feedback-limited stellar mass following the analytic arguments of \citet{grudic_does_2023}. For each star-forming group, we define the parent cloud using gas with $n_{\rm H} \geq 10^2 \; \rm cm^{-3}$ and adopt $R_{90}$, the radius enclosing $90 \%$ of the selected cloud mass, as the characteristic cloud radius. The cloud surface density is estimated as $\Sigma = \frac{\rm M_{\rm cl}}{\pi \rm R^2_{90}}$. For photoionization-regulated cloud disruption, they estimate the maximum stellar mass in the cloud to be 


\begin{equation}\label{eq:mass scaling 1 (photodissociation)}
\begin{split}
M_{\star,\max}
\gtrsim\;&
16\,M_\odot\,
N_{\rm ff}^{-0.55}
\left(
\frac{\Sigma}
{100\,M_\odot\,{\rm pc}^{-2}}
\right)^{0.34} \\
&\times
\left(
\frac{M_{\rm cl}}
{10^4\,M_\odot}
\right)^{0.24}
\left(
\frac{T_i}
{10^4\,{\rm K}}
\right)^{-0.39},
\end{split}
\end{equation}

where $N_{\rm ff}$ is the disruption time in units of the cloud free-fall time and $T_{i}$ is the temperature of the ionized gas. This analytic model assumes that the ionizing radiation is dominated by the most massive main-sequence star in the cloud. Adopting $N_{\rm ff}=1$ and $T_i=2\times10^4\,{\rm K}$ as an approximate value for ionized metal-free gas, this relation gives characteristic maximum stellar masses of $\sim14$, $6$, $10$, and $7\,M_\odot$ for star-forming Minihalos 2--5, respectively. These values are broadly comparable to the maximum stellar masses measured in the simulation in each of those minihalos, $10.0$, $3.9$, $4.7$, and $8.4\,\rm M_\odot$, suggesting that photoionization feedback may contribute to limiting continued stellar growth in these lower-mass systems.

The first star-forming group in the simulation is different. Its parent cloud is the most massive and has the highest surface density in our sample, with $M_{\rm cl}\approx8.2\times10^4\,M_\odot$, $R_{90}\approx11.4\,{\rm pc}$, and $\Sigma\approx2.0\times10^2\,M_\odot\,{\rm pc}^{-2}$. This cloud forms the most massive star in the simulation, with $M_{\star,\max}\approx110\,M_\odot$. Because the most massive stars in our simulation are sufficiently massive for their luminosities to approach the Eddington-limited scaling, we adopt the Eddington-limited branch of the momentum-regulated relation from \citet{grudic_does_2023},


\begin{equation}\label{eq:mass scaling 2 (momentum driven)}
M_{\star,\max}
=
100\,M_\odot
\frac{
\left(
\dfrac{M_{\rm cl}}
{10^5\,M_\odot}
\right)
\left(
\dfrac{\Sigma}
{100\,M_\odot\,{\rm pc}^{-2}}
\right)
}{
(1-f_{\rm esc})N_{\rm ff}^{2}
}.
\end{equation}

Adopting $f_{\rm esc}=0$ and $N_{\rm ff}=1$, consistent with the order-of-magnitude assumption $(1-f_{\rm esc})N_{\rm ff}^2\sim1$ adopted by \citet{grudic_does_2023} for the Eddington-limited branch, gives $M_{\star,\max}\sim164\,M_\odot$, the same order of magnitude as the $\sim110\,M_\odot$ star formed in Minihalo 1. We also check whether the stellar population in Minihalo 1 is large enough to sample the high-mass end of its IMF. From Figure \ref{fig:IMF for each different halo}, Minihalo 1 has a high-mass slope of $dN/dM\propto M^{-1}$. For a power law truncated at $M_{\max}$, the expected number of stars above a threshold $M_{\rm th}$ is approximately

\begin{equation}\label{eq:mass threshold from imf}
N(>M_{\rm th})\sim
\frac{M_{\star,\rm tot}}{M_{\max}}
\ln\left(\frac{M_{\max}}{M_{\rm th}}\right).
\end{equation}

Using $M_{\star,\rm tot}\simeq320\,M_\odot$, \(M_{\max}\simeq110\,M_\odot\), and $M_{\rm th}=100\,M_\odot$, we find $N(>100\,M_\odot)\sim0.3$. Thus, the formation of a $\sim110\,M_\odot$ star in Minihalo 1 is plausible despite the relatively small stellar population.

The mass and surface density of the first cloud therefore allow the most massive star to remain connected to a large, dense gas reservoir and continue accreting to much higher mass before feedback becomes capable of disrupting or removing that reservoir. Applying Equation \ref{eq:mass scaling 2 (momentum driven)} to the other star-forming clouds gives lower limiting stellar masses than for the first group of stars to form. Thus, both the photoionization-regulated (Equation \ref{eq:mass scaling 1 (photodissociation)}) and momentum-regulated (Equation \ref{eq:mass scaling 2 (momentum driven)}) estimates are consistent with the first group of stars producing a larger maximum stellar mass than the later-forming groups. These analytic estimates are approximate and depend on the adopted cloud properties and parameters such as \(N_{\rm ff}\), \(T_i\), and \(f_{\rm esc}\), so we use them only for comparison with the simulated stellar masses. These estimates should therefore be interpreted as order-of-magnitude estimates rather than precise predictions, but still provide valuable intuition. 

\subsubsection{Protostellar and Radiative Feedback}
The role of protostellar feedback is especially important in this simulation because jets can regulate stellar growth in more than one way. Our previous work \citep{meziani_forged_2026}, looking at the evolution of one Pop.~III star using FORGE'd in FIRE, found that protostellar jets suppress accretion onto the star not only by ejecting mass from the star, but also by injecting momentum into the surrounding gas. This momentum loading can prevent additional gas from reaching the (proto)star, so the total reduction in accretion onto the star can be larger than the mass directly ejected by the jets. This was also found to be true in the case of Pop.~I star formation in idealized simulations of molecular clouds in MW-like environments at $z=0$ \citep{guszejnov_effects_2022}. Thus, even if the most massive stars continue accreting at high rates, protostellar jets can still limit their final masses by reducing the gas supply available for continued growth. In addition to jets, radiative feedback from massive stars can heat and dissociate the surrounding gas through UV, Lyman-Werner, and IR continuum radiation. This reduces the amount of cold, dense gas available for continued growth and can therefore limit competitive accretion.


\subsubsection{Dynamical Ejections}
Dynamical ejections provide another mechanism for limiting the time that (proto)stars spend accreting gas. As discussed in Section \ref{subsec:stellar ejection}, $16$ of the $56$ (proto)stars are classified as unbound (see Figure \ref{fig:ejections}). If these stars become unbound while they are still accreting, they no longer have access to the dense gas reservoir from which they were accreting, effectively limiting their final masses. These ejections are likely driven by dynamical interactions between multiple stars. Most of the binary and higher-order multiple systems in our simulation form in the first minihalo, which is also where the majority of the low-mass stars originate (see Figure \ref{fig:IMF for each different halo}, Minihalo 1). This suggests that dynamical interactions in the first minihalo could contribute to some of the unbound stars identified in Figure \ref{fig:ejections}. 

\subsubsection{Supernova Feedback and Continued Star Formation}

At around $z \sim 11.79$, a second group of Pop. III stars forms in a different minihalo, which we refer to as Minihalo 2. Figure \ref{fig:velocity and DM plots} shows the system following supernova feedback from this group, with the projected gas velocity map (top-left) and dark matter surface density (top-right). The gas velocity map shows strong outflows that extend beyond Minihalo 2, indicating that the supernovae have expelled gas from the minihalo. The bottom panels show a closer view of Minihalo 1 (bottom-left) and Minihalo 2 (bottom-right). The dashed white circles mark the $R_{200m}$ of each minihalo, with some stars located beyond these radii. Although supernova feedback suppresses further star formation within Minihalo 2, Pop. III stars continue to form in other minihalos until the end of the simulation.

These results suggest that several processes regulate Pop. III star formation. Protostellar jets and ionizing radiation limit the growth of individual stars by reducing the gas available for accretion. Supernova feedback, on the other hand, suppresses subsequent star formation within the host minihalos. This is particularly important for the relatively small minihalos in our simulation, where the low escape velocities allow gas to be expelled more easily.

As discussed above, the star-forming clouds in our simulation contain only a few thousand to tens of thousands of solar masses of gas and have relatively modest surface densities (of order $\sim 100 \; \rm M_{\odot}/pc^2$). Previous work has shown that the mass and surface density of star-forming clouds can influence the formation and maximum masses of massive stars \citep{guszejnov_evolution_2020, grudic_does_2023}. Therefore, the high-mass cutoff seen in Figure \ref{fig:IMF cumulative} likely reflects the combined effects of stellar feedback and the limited mass and surface density of the gas reservoirs from which these stars form.

\subsection{Comparison to Previous IMF Studies}

In this work, we do not isolate the dominant processes by turning individual physics channels on and off. However, previous STARFORGE studies have already explored how feedback, cloud properties, and low-metallicity gas shape the IMF. In particular, \citet{guszejnov_effects_2022} found that the IMF is relatively insensitive to several cloud-scale initial conditions, including the initial turbulence, cloud mass, and cloud surface density, even though these parameters strongly affect the star formation history and final star formation efficiency. Instead, they found that the characteristic stellar mass depends more directly on physics that changes the gas temperature, including metallicity and the interstellar radiation field. They also showed that radiative feedback plays an important role in quenching star formation and disrupting the parent cloud, while protostellar jets are especially important for setting the IMF peak. Their low-metallicity simulations produce a shallower high-mass IMF than the Kroupa IMF. In a separate STARFORGE study, \citet{grudic_dynamics_2022} found a Chabrier-like IMF with a high-mass (M $> 0.5 \; \rm M_{\odot}$) slope of $\alpha \sim 2$ when the combined effects of protostellar jets, radiation, stellar winds, and supernova feedback are included. The simulation analyzed here extends this question to the metal-free regime and finds a similarly shallow, but not extremely top-heavy, high-mass IMF.

Previous Pop.~III simulations provide an important comparison because they showed that primordial star formation is neither purely isolated nor restricted to extremely massive stars. \citet{susa_mass_2014} used cosmological simulations to identify 59 minihalos, which were then resimulated with local three-dimensional radiation-hydrodynamic calculations, and found stellar masses spanning $1 \; \rm M_{\odot} \lesssim M_{\star} \lesssim 300 \; \rm M_{\odot}$ across 59 minihalos, with the distribution peaking at several $\times 10 \; \rm M_{\odot}$ and about two-thirds of the minihalos forming multiple systems. \citet{stacy_building_2016} followed one Pop.~III system from cosmological initial conditions to much higher spatial resolution, resolving scales of $\sim 1 \; \rm au$, and found 37 sink particles with masses spanning $0.02$--$43 \; \rm M_{\odot}$. Their IMF remained top-heavy relative to the present-day IMF ($\alpha \sim 1.2$), with most of the stellar mass contained in stars above $\sim 1 \; \rm M_{\odot}$. More recently, \citet{latif_birth_2022} performed cosmological radiation-hydrodynamic simulations of five primordial halos and found 55 stars with masses of $0.1$--$40 \; \rm M_{\odot}$ and an IMF slope of $dN/dM \propto M^{-1.2}$, with $\sim 70\%$ of stars ejected from their disks. Compared to the shallower high-mass IMF slopes of $\alpha\sim1.2$ found by \citet{stacy_building_2016} and \citet{latif_birth_2022}, our simulation produces a steeper slope of $\alpha=1.9$. These differences may partly reflect the additional physics included in FORGE'd in FIRE, particularly magnetic fields and protostellar jets, which are not included together in these previous simulations. These simulations followed protostellar evolution over different timescales, ranging from $\sim 5000 \; \rm yr$ in \citet{stacy_building_2016} to $\sim 10^{5} \; \rm yr$  in \citet{susa_mass_2014}, while \citet{latif_birth_2022} evolved their halos for up to $\sim 20 \; \rm kyr$ after the first star formed. Our simulation follows star formation over approximately $500$ Myr, allowing us to capture continued star formation and feedback across multiple minihalos.




As shown in Figure \ref{fig:IMF over time}, the early-time PDMF in this simulation is broadly consistent with the Pop.~III simulations discussed above. In the top two panels, corresponding to approximately the first $\lesssim 1 \; \rm Myr$ after the first sink forms, the high-mass slope is $\alpha \sim 1.4 - 1.5$, similar to the slopes found in previous Pop.~III studies. The turnover mass remains at $M_{\rm turn} \approx 0.5 \; \rm M_{\odot}$ over these early snapshots. By $z=7.27$, approximately $490$ Myr after the first stars form, the turnover mass increases to $M_{\rm turn} \approx 1.5 \; \rm M_{\odot}$.
By approximately $8$ Myr after the first stars form, the PDMF slope has steepened to $\alpha \sim 1.8$, and by $\sim 490$ Myr it reaches $\alpha \sim 2.6$. Thus, the final IMF analyzed in this paper (Section \ref{subsec:initial mass function}), which includes all unique stars formed throughout the simulation, likely reflects the longer-term evolution of the stellar population under the additional accretion-regulating physics included in this simulation, including RMHD, magnetic fields, protostellar jets and winds, radiation, supernova feedback, and detailed primordial cooling and chemistry. The comparison therefore suggests that the IMF shown in Figure \ref{fig:IMF cumulative} (top panel) does not necessarily contradict earlier Pop.~III calculations: at comparable early evolutionary times, our simulation produces a similarly top-heavy stellar mass distribution, while continued star formation and feedback over later times produce a steeper PDMF and contribute to the final IMF measured over the full stellar population. 

\section{\label{sec:conclusions} Summary and Conclusions}

We have presented a cosmological RMHD simulation of Population III star formation beginning at \(z\sim16\) in a dark matter halo with a mass of \(M_{\rm halo}\sim2.4\times10^6\,M_\odot\), whose descendant reaches $M_{\rm halo}\sim2\times10^9\,M_\odot$ at $z=0$, using the FORGE’d in FIRE framework. This extends our previous single-star study to a full stellar population, allowing us to characterize the IMF and multiplicity in a self-consistent cosmological environment. The simulation follows Pop.~III star formation within an ultra-faint dwarf progenitor for approximately $500$ Myr and includes magnetic fields, radiation transport, primordial chemistry and cooling, protostellar jets and winds, and supernova feedback. 

Star formation proceeds in discrete bursts, producing 56 stars across five distinct star-forming minihalos of masses in the range $M_{200 \rm m} \sim 10^6 - 10^7\,\rm{M_\odot}$. The first star-forming halo converts only a small fraction of its estimated total baryonic mass, inferred from the cosmic baryon fraction, into stars, with an SFE of $\epsilon_{\rm SFE} \approx 6.8 \times 10^{-4}$. The population is dominated by single systems, with a system-based multiplicity fraction of $\approx0.15$, while approximately $29\%$ of individual stars belong to multiple systems. 


The IMF is only mildly top-heavy relative to the Kroupa IMF at the resolved high-mass end, with $\alpha = 1.9$. However, this slope is similar to that found in STARFORGE simulations of MW-like molecular clouds. \citet{grudic_dynamics_2022} find a high-mass slope of $\alpha \sim 2.0$ with the full STARFORGE feedback model applied to an analog of a star-forming region in the MW at the present day, which is also somewhat shallower than the canonical Kroupa/Salpeter slope. Thus, the high-mass slope measured here is not significantly different from the current STARFORGE prediction under MW-like conditions at $z=0$ and should not be interpreted as a unique consequence of metal-free star formation.

The stellar mass distribution also evolves substantially with time. At early times, the present-day mass function has a high-mass slope of $\alpha \sim 1.4-1.5$, comparable to previous Pop.~III simulations, whereas the PDMF steepens as the simulation evolves. The IMF measured using all unique stars formed throughout the simulation has a final high-mass slope of $\alpha \sim 1.9$, as shown in Figure \ref{fig:IMF cumulative} (top panel). This suggests that differences between our final IMF and previous Pop. III calculations may partly reflect the much longer simulation evolution. The maximum stellar mass also varies between star-forming halos. The first and most massive parent cloud produces the $\sim110\, \rm M_\odot$ star, while later, lower-mass clouds form stars with maximum masses of only $\sim4-10\,M_\odot$. Analytic feedback-limited estimates based on the cloud masses and surface densities are broadly consistent with these maximum stellar masses, suggesting that the available gas reservoir and feedback both regulate stellar growth.

We find that $16$ of the $56$ stars (\(\sim29\%\)) are classified as unbound. Three of these stars have masses below \(1\,M_\odot\), including objects sufficiently low in mass to potentially survive to the present day. This provides a possible pathway for primordial low-mass stars to leave their early star-forming environments and later become incorporated into larger galaxies. Recent work by \citet{ferreira_red-giant_2026} suggests that low-mass Pop.~III stars surviving to $z=0$ could potentially be identified through their distinct asteroseismic signatures, even if their surfaces have been enriched by subsequent accretion.

There are many ways that we can improve our current simulation, and it is important for us to mention these limitations. First, the way that we deal with binaries could be improved, particularly during the merging process. Instead of merging the stars into one sink and considering the merged system as a single star, we could have several different \enquote{tracks} that allow us to consider each Pop. III star separately and follow their individual properties over time after merging. Another improvement would be to include a Lyman-Werner background, which we currently do not model in our simulation. Such a background could suppress H$_2$ cooling in minihalos before they form stars, delaying gravitational collapse until they reach higher halo masses and potentially changing the timing and distribution of Pop.~III star formation \citep{machacek_simulations_2001, oshea_population_2008, safranek-shrader_star_2012, brown_lyra_2026}.  

It would also be useful to follow these simulations in different environments instead of only studying a UFD over the timescale considered here. This could include trying different environments and different galaxy initial conditions. Finally, we include protostellar jets in this simulation, but whether Pop. III stars actually launch protostellar jets is still uncertain \citep{machida_first_2006, machida_magnetohydrodynamics_2008, prole_primordial_2022, meziani_forged_2026}. Future simulations could therefore explore how the stellar population changes when protostellar jet feedback is varied or excluded.

In a follow-up paper, we will study the supernova ejecta from this simulation, where the ejecta travel, and how they enrich the surrounding gas. We also plan to improve the stellar-remnant model so that white dwarfs, neutron stars, and black holes are followed more realistically.

\begin{acknowledgments}
The authors would like to thank Stella S. R. Offner for useful discussions. Numerical calculations and plots in this paper were run on the Texas Advanced Computing Center (TACC) allocations AST21010 and TG-PHY260478.
Support for P.F.H. was provided by a Simons Investigator grant. 
C.A.F.G. was supported by NSF through grants AST-2108230, AST-2307327 and AST-2606015; by NASA through grants 80NSSC22k0809, 80NSSC22K1124 and 80NSSC24K1224; by STScI through grant JWST-AR-03252.001-A; and by BSF through grant \#2024262. ChatGPT (GPT-6) was used in the editing and revision of this paper's text and figure formatting.
\end{acknowledgments}


\bibliography{references}

\end{document}